\documentclass[aps,prd,10pt,onecolumn,floatfix,superscriptaddress,preprintnumbers,shownopacs,showkeys,nofootinbib,notoccite,notitlepage,preprintnumbers]{revtex4-2}

\usepackage{graphicx}
\usepackage{dcolumn}
\usepackage{bm}
\usepackage[colorlinks = true, linkcolor = purple, urlcolor  = blue, citecolor = blue, anchorcolor = blue]{hyperref}
\usepackage{braket}
\usepackage{color}
\usepackage{amsmath}
\usepackage{float}
\usepackage{ulem}
\usepackage{subfig}
\usepackage{ragged2e}

\usepackage{tikz,xcolor,hyperref}
\definecolor{darkgreen}{rgb}{0.0, 0.2, 0.13}
\definecolor{bostonuniversityred}{rgb}{0.8, 0.0, 0.0}
\definecolor{lime}{HTML}{A6CE39}
\DeclareRobustCommand{\orcidicon}{
	\begin{tikzpicture}
	\draw[lime, fill=lime] (0,0) 
	circle [radius=0.16] 
	node[white] {{\fontfamily{qag}\selectfont \tiny ID}};
	\draw[white, fill=white] (-0.0625,0.095) 
	circle [radius=0.007];
	\end{tikzpicture}
	\hspace{-2mm}
}
\foreach \x in {A, ..., Z}{\expandafter\xdef\csname orcid\x\endcsname{\noexpand\href{https://orcid.org/\csname orcidauthor\x\endcsname}
			{\noexpand\orcidicon}}
}
\begin{document}
\preprint{FERMILAB-PUB-25-0620-T}
\title{Cosmic-ray boosted inelastic dark matter from neutrino-emitting active galactic nuclei}

\author{R. Andrew Gustafson\hspace{-1mm}\orcidA{}}
\email{gustafr@vt.edu}
\affiliation{Center for Neutrino Physics, Department of Physics, Virginia Tech, Blacksburg, Virginia 24061, USA}
\affiliation{International Center for Quantum-field Measurement Systems for Studies of the Universe and Particles (QUP,WPI), High Energy Accelerator Research Organization (KEK), Oho 1-1, Tsukuba, Ibaraki 305-081, Japan}

\author{Gonzalo Herrera\hspace{-1mm}\orcidB{}}
\email{gonzaloh@mit.edu}
\affiliation{Center for Neutrino Physics, Department of Physics, Virginia Tech, Blacksburg, Virginia 24061, USA}
\affiliation{Kavli Institute for Astrophysics and Space Research, Massachusetts Institute of Technology, Cambridge, MA 02139, USA}
\affiliation{Harvard University, Department of Physics and Laboratory for Particle Physics and Cosmology, Cambridge, MA 02138, USA}

\author{Mainak Mukhopadhyay\hspace{-1mm}\orcidC{}}
\email{mainak@fnal.gov}
\affiliation{Department of Physics; Department of Astronomy \& Astrophysics; Center for Multimessenger Astrophysics, Institute for Gravitation and the cosmos, The Pennsylvania State University, University Park, Pennsylvania 16802, USA}
\affiliation{Astrophysics Theory Department, Theory Division, Fermi National Accelerator Laboratory, Batavia, Illinois 60510, USA}
\affiliation{Kavli Institute for Cosmological Physics, University of Chicago, Chicago, Illinois 60637, USA}

\author{Kohta Murase\hspace{-1mm}\orcidD{}}
\email{murase@psu.edu}
\affiliation{Department of Physics; Department of Astronomy \& Astrophysics; Center for Multimessenger Astrophysics, Institute for Gravitation and the cosmos, The Pennsylvania State University, University Park, Pennsylvania 16802, USA}
\affiliation{Center for Gravitational Physics and Quantum Information, Yukawa Institute for Theoretical Physics, Kyoto, Kyoto 606-8502 Japan}

\author{Ian M. Shoemaker\hspace{-1mm}\orcidE{}}
\email{shoemaker@vt.edu}
\affiliation{Center for Neutrino Physics, Department of Physics, Virginia Tech, Blacksburg, Virginia 24061, USA}

\begin{abstract}
Cosmic rays may scatter off dark matter particles in active galactic nuclei, where both the densities of cosmic rays and dark matter are expected to be very large. These scatterings could yield a flux of boosted dark matter particles directly detectable on Earth, which enhances the sensitivity of dark matter direct detection and neutrino experiments to light and inelastic dark matter models. Here we calculate the cosmic-ray boosted dark matter flux from the neutrino-emitting active galactic nuclei, NGC 1068 and TXS 0506+056, by considering realistic cosmic-ray distributions, deep inelastic scatterings, and mass splittings in the dark sector. From this we derive novel bounds from these sources on light and/or inelastic dark matter models with Super-K and XENONnT. We find that cosmic-ray boosted dark matter from neutrino-emitting active galactic nuclei can test regions of parameter space favored to reproduce the observed relic abundance of dark matter in the Universe, and that are otherwise experimentally inaccessible.
\end{abstract}

\maketitle

\section{\label{sec:introduction}
Introduction}

Evidence for gravitational effects of non-luminous matter at different cosmological scales has been established through multiple observations \cite{Bertone:2004pz, Cirelli:2024ssz}. A well-motivated paradigm consists of dark matter (DM) being composed of new particles with weak or feeble interactions with the Standard Model (SM). A plethora of experiments have been devoted to search for DM particles from the Galactic halo via their elastic scatterings off nuclei and/or electrons at Earth-based detectors \cite{Goodman:1984dc, Bernabei:2007gr, Essig:2011nj, Essig:2022dfa}. As of today, no conclusive signal has been found, which allows us to set constraints of DM particles with masses from the MeV to the TeV scale \cite{LZ:2022lsv, SENSEI:2023zdf, DAMIC-M:2025luv, Zhang:2025ajc}.

Constraints on spin-independent interactions in the mass range from 1 GeV to 1 TeV restricts several DM scenarios able to address the electroweak hierarchy problem while yielding the observed relic abundance of the Universe via freeze-out \cite{Jungman:1995df}. The constraints on MeV scale DM are still orders of magnitude weaker than for GeV-TeV scale DM, but several thermal and nonthermal production scenarios are already probed with some direct detection experiments, especially in light of recent results from the DAMIC-M and PANDAX-4T experiments \cite{DAMIC-M:2025luv, Krnjaic:2025noj, PandaX:2024muv, Cheek:2025nul}.

There is an important exception which is poorly tested by direct detection searches. In some DM models, the inelastic scattering channel can naturally dominate over the elastic one. A canonical example is the vector current of a Majorana particle, which is forbidden for the elastic case, but not for the inelastic one. This situation has a close analogue in neutrino physics. For Majorana neutrinos, the diagonal magnetic moment operator vanishes identically by CPT invariance \cite{Nieves:1981zt,Kayser:1982br,Giunti:2014ixa}, so the elastic channel $\nu_i \to \nu_i$ is absent. Only off-diagonal magnetic moments are allowed, which mediate inelastic processes of the form $\nu_i \to \nu_j$ with $i\neq j$. A similar mechanism appears in the Minimal Supersymmetric Standard Model, where the lightest supersymmetric particle can be dominantly Higgsino and much lighter than the other supersymmetric states. In that case, the elastic scattering channel is suppressed by the large masses of the supersymmetric partners, and the inelastic scattering channel induced by the electroweak gauge interactions can dominate \cite{Nagata:2014wma}.

Beyond these examples, several realizations of inelastic DM with mass splittings ranging from $\sim$ eV to $\sim$ MeV have been discussed in the literature, showing that simplified models can account for the DM of the Universe, and that a variety of phenomenological probes can constrain these models, {\it e.g.} in  \cite{Hall_1998,Alves_2010,Schwetz_2011,Arkani_Hamed_2009,McCullough_2010,Chang_2010,Barello_2014, Nagata:2014wma, Nagata_2015_2, Fujiwara:2022uiq, Chatterjee:2022gbo,Herrera:2023fpq, Berlin:2023qco, Chauhan:2023zuf,Garcia:2024uwf, DallaValleGarcia:2024zva,Gustafson:2024aom,Berlin:2025fwx}. Such inelastic models are interesting, as the scattering is suboptimal for direct detection probes.

It has been discussed in a variety of works that models of light and/or inelastic DM can be probed when considering scatterings of cosmic rays(CRs) off DM in the Milky Way, which would yield a boosted flux of DM particles on Earth that could be detected with DM and neutrino experiments, see \textit{e.g} \cite{Bringmann:2018cvk, Ema:2018bih, Cappiello:2024acu,Su:2023zgr,Liang:2024xcx,Feng:2021hyz,Bell:2021xff}, and the scattering could also lead to a flux of secondary gamma-rays and neutrinos \cite{Guo:2020oum,Bell:2021xff}. These works were only able to probe relatively large scattering cross sections of DM off protons and electrons, coinciding with regions of parameter space that are mostly ruled out by several other probes. However, it was demonstrated in \cite{Wang:2021jic} that a larger boosted DM flux on Earth could instead arise from some blazars, due to the large CR proton flux and expected DM density in these environments.

In this work, we calculate the flux of CR boosted DM reaching the Earth from NGC 1068 (a Type-II Seyfert galaxy) and TXS 0506+056 (a blazar), two well known multi-messenger sources and CR accelerators. We derive upper limits on a simplified model of DM from a non-observation of an excess of events induced by DM interactions with nuclei and electrons at Earth-based experiment Super-Kamiokande (Super-K) \cite{Super-Kamiokande:2022ncz} (and XENONnT \cite{XENON:2023cxc} in Appendix \ref{sec:nuclear_recoils_Xenon}). We discuss the complementarity of our constraints with those obtained from other searches, demonstrating that active galactic nuclei (AGN) allow us to improve over previous bounds in a large region of parameter space. Previous works considering scatterings of DM particles off CRs in AGN focused either on spectral distortions on the observed gamma-ray fluxes \cite{Gorchtein:2010xa,Ambrosone:2022mvk}, CR electron scatterings off DM \cite{Xia:2024ryt,Herbermann:2024kcy}, the implications of CR-DM interactions for multi-messenger observations including high-energy neutrinos \cite{Herrera:2023nww, Gustafson:2024aom, DeMarchi:2024riu, DeMarchi:2025xag, Wang:2025ztb, DeMarchi:2025uoo}, or studied the boosted DM flux considering only the multi-messenger blazar TXS 0506+056 \cite{Wang:2021jic, Granelli:2022ysi, Bhowmick:2022zkj, Jeesun:2025gzt, DeMarchi:2024riu} or considering additional blazars that do not present evidence for CR proton acceleration \cite{DeMarchi:2025uoo, DeMarchi:2025uoo}. These works generally present predictions without addressing CR model uncertainties, and did not consider inelastic DM scenarios. In this work we address all these tasks in detail for TXS 0506+056, further computing the boosted DM flux from the steady neutrino-emitting source NGC 1068.

The paper is organized as follows. In Section \ref{sec:AGN} we describe our modeled CR flux from NGC 1068 and TXS 0506+056, discussing the associated uncertainties. In Section~\ref{sec:dm_agn} we describe the DM density profile and column density in NGC 1068 and TXS 0506+056. In Section \ref{sec:Boostedfluxes} we compute the CR boosted DM flux from NGC 1068 and TXS 0506+056, considering deep inelastic scattering (DIS) and mass splittings in the dark sector. We also comment on the attenuation of a DM flux as it passes through Earth. In Section \ref{sec:Rates} we discuss the scattering signatures of this flux at Super-K, and derive upper limits on the DM parameters. We further confront these limits with complementary constraints and motivated regions of parameter space able to reproduce the relic abundance of the Universe. Finally, in Section \ref{sec:conclusions}, we present our conclusions.

\section{Cosmic-ray flux in active galactic nuclei \label{sec:AGN}}
Recent high-energy neutrino observations from TXS 0506+056~\cite{IceCube:2018cha,IceCube:2018dnn,IceCube:2023oua} and NGC 1068 \cite{IceCube:2022der}, together with the EM counterparts, which are produced via $pp$, $p\gamma$ interactions, plus leptonic processes such as inverse compton scattering or synchroton radiation, allow us to infer the energy budget of CRs. The so-called leptohadronic models have been able to explain the neutrino and EM spectra simultaneously under certain conditions, and the accelerating region of CRs has been constrained to some extent~\cite{Cerruti:2018tmc,Murase:2022dog}. Below we discuss the details of the CR proton spectra in these sources.
%
\begin{figure*}
\centering
\includegraphics[width=0.49\linewidth]{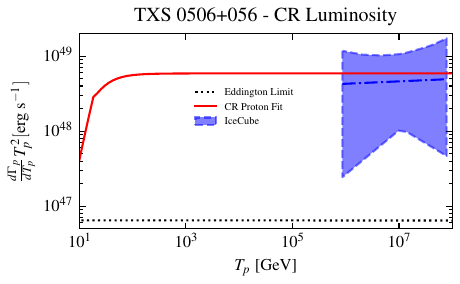}
\includegraphics[width=0.49\linewidth]{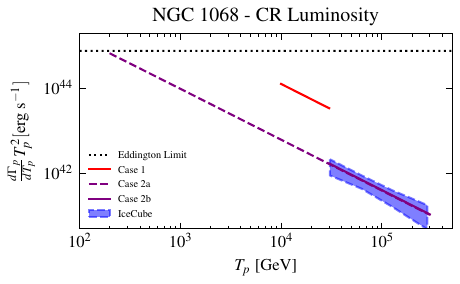}
\caption{\justifying \label{fig:ngc1068_crlum} The CR luminosity for TXS 0506+056 (\textit{left}) and NGC 1068 (\textit{right}). See Table~\ref{fig:table1} for details on the parameters and Eq. \ref{eq-AGN-Flux} for the explicit calculation. The results from the IceCube Collaboration~\cite{IceCube:2022der} are shown as the blue band with the best fit line shown as a dash-dot line. For NGC 1068, the cases (C1 and C2) are discussed in Sec.~\ref{subsec:ngccrspec} and  the results from the IceCube collaboration~\cite{IceCube:2022der} are shown as the blue band with the best fit line represented in solid dark blue.
}
\end{figure*}

\begin{table*}
    \centering
    \begin{tabular}{|>{\centering\arraybackslash}m{3cm}|>{\centering\arraybackslash}m{3cm}|>{\centering\arraybackslash}m{3cm}|>
    {\centering\arraybackslash}m{5cm}|}
        \hline
        \textbf{Parameter} & \textbf{TXS 0506+056} & \textbf{NGC 1068: C1} & \textbf{NGC 1068 : C2a [C2b]} \\
        \hline
        $d_{L} (\mathrm{Mpc})$ & 1750 & 10 & 10 \\
        $M_{\mathrm{BH}} (M_{\odot})$ &  $5 \times 10^8$ & $6 \times 10^6$ & $6 \times 10^6$\\
        $\Gamma_B$ & 20 & 1 & 1\\
        $\theta_{\mathrm{LOS}} (^{\circ})$ & 0 & 0 & 0\\
        $\alpha_p$ & 2.0 & 3.2 & 3.2\\
        $\gamma^\prime_{\min, p}$ & 1.0 & $10^{4}$ & $1.4 \times 10^{2}$ [$3 \times 10^{4}$]\\
        $\gamma^\prime_{\max, p}$ & $1.6 \times 10^7$ & $3 \times 10^4$ & $3 \times 10^5$\\
        $L_p (\mathrm{erg} / \mathrm{s})$ & $8.6 \times 10^{49}$ & $7 \times 10^{43}(\sim L_X)$ & $7.8 \times 10^{44} (\sim L_{\rm edd})$ [$1.2 \times 10^{42}$]\\
        $R_{\rm acc} (R_S)$ & $1.2 \times 10^5$ & $30$ & $30$\\
        $r_0 (\mathrm{kpc})$ & 10 & 10 & 10\\
        \hline
    \end{tabular}
    \caption{ \justifying Parameters considered in this work, needed to compute the CR proton spectra and DM distribution on NGC 1068 and TXS 0506+056. For NGC 1068 the corresponding values for the Case C2b are written within [.] if different from Case C2a. The minimum/maximum proton boost factor \textit{in the frame of the blob} $\gamma^{\prime}_{\min/\max,p}$ is provided, along with the spectral index of the cosmic rays \textit{in the frame of the blob}, $\alpha_{p}$. The variable $\theta_{\rm{LOS}}$ is the angle between the jet and line-of-sight, here assumed to be 0 in all cases (i.e. either there is no jet, or it is directed at Earth). The radius of cosmic ray acceleration $R_{acc}$ and the scale radius of the host galaxy $r_{0}$ are both provided, which are important in calculating the integrated dark matter density through which these cosmic rays pass.} 
    \label{fig:table1}
\end{table*}
\subsection{TXS 0506+056}
%
%
%
TXS 0506+056 is a blazar (an AGN with a relativistic jet pointing towards the observer) at redshift $z\approx 0.33$ ($d_L = 1750$ Mpc) from which a neutrino event (IceCube-170922A) of energy $0.1 - 1$ PeV was reported~\cite{IceCube:2018dnn} at the level of $3 \sigma$ while there was an ongoing gamma-ray flare. The neutrino observation was followed up by various electromagnetic (EM) observations in the high-energy and very high-energy gamma-rays, hard and soft X-rays, ultraviolet, optical, and near-infrared wavelengths~\cite{Keivani:2018rnh} (see references therein). In addition to this, an archival search by the IceCube Collaboration at the location of TXS 0506+056 lead to significant association of another $13 \pm 5$ neutrino events~\cite{IceCube:2018cha}. The presence of such high energy signatures both in the EM and neutrino channels make TXS 0506+056 an ideal site of particle acceleration~\cite{Murase:2018iyl}.

The spectral energy distribution of TXS 0506+056 from optical to gamma-rays can be explained by accelerated primary electrons that undergo inverse Compton and synchrotron processes. In fact, to simultaneously explain the high-energy neutrino and gamma-ray observations such leptonic models are amongst the only viable ones, where the radiation from accelerated protons are hidden due to EM cascades in the source and $\gamma \gamma$-attenuation. The neutrino and proton luminosities are thus strongly constrained by X-ray observations, else the high energy photons produced from photohadronic interactions would violate the X-ray data.

The differential CR proton luminosity is related to the total luminosity by
\begin{align}
\label{eq:crlum}
L_p & = \int_{T_p^{\rm min}}^{T_p^{\rm max}}\ d T_p\ (T_p + m_{p}) \frac{d \Gamma_{p}}{dT_{p}}\,,
\end{align}
where $T_{p}$ is the CR proton energy kinetic in the source frame, $\Gamma_{p}$ is the production rate of protons, $L_p$ is the normalization, and $T_p^{\rm min}$ and $T_p^{\rm max}$ give the minimum and maximum CR kinetic energies respectively. In the case with no jet (that is $\Gamma_{B} = 1$), then $T_{p}^{\min/\max} = m_{p} \gamma^{\prime}_{\min/\max,p}$, where $\gamma^{\prime}_{\min/\max,p}$ is the minimum/maximum boost factor in the frame of the blob. In the case with a jet, $\gamma^{\prime}_{\min/\max,p}$ are still used as cutoffs in the blob frame, and the velocity distribution is boosted to obtain $T^{\min/\max}_{p}$ in the lab frame. In Fig.~\ref{fig:ngc1068_crlum} we show the CR proton spectra in terms of $T_p^2 \frac{d \Gamma_{p}}{dT_{p}}$. We show the results from the IceCube Collaboration~\cite{IceCube:2018cha} in blue. For TXS 0506+056, based on IceCube observations, the CR protons are accelerated from 0.2 PeV to 2 PeV. 
The best fit line along with the upper and lower bounds are obtained by converting the neutrino observations assuming $T_p \approx 20 T_\nu$. The differential CR proton spectrum is approximately related to the neutrino spectrum by~\cite{Murase:2018iyl,Murase:2022dog}
\begin{equation}
\label{eq:diffptonu}
T_p \frac{d \Gamma_{p}}{dT_{p}} \approx \frac{4(1+K)}{3K} f_{p\gamma}\  T_\nu \frac{d \Gamma_{\nu}}{dT_{\nu}}\,,
\end{equation}
where we have $K=1 (2)$ $p\gamma$ ($pp$) processes, $f_{p\gamma}$ is defined as the efficiency of pion production, that is, the fraction of CR protons producing pions through the
photomeson process. Observations of gamma-rays between 10 - 100 GeV imply the inefficiency of neutrino production in the same region. Thus setting the optical depth to be less than 1 at 100 GeV implies $f_{p\gamma} \sim 10^{-4}$~\cite{Murase:2018iyl} (see Sec. 3 there). The Eddington luminosity assuming a black hole mass $M_{\rm BH} \sim 5 \times 10^8 M_\odot$ is shown with a black dotted line\footnote{The Eddington luminosity ($L_{\rm Edd}$) is obtained by balancing the outward radiation pressure and the inward gravitational force. In hydrostatic equilibrium
\begin{equation}
L_{\rm Edd}
= \frac{4\pi G c\, m_p}{\sigma_T}\, M_{\rm BH}
\simeq 1.3\times 10^{45}\ {\rm erg\,s^{-1}}
\left(\frac{M_{\rm BH}}{10^{7} M_\odot}\right),
\label{eq:Ledd}
\end{equation}
where $M_{\rm BH}$ is the mass of the black hole and $
\sigma_T = 6.65 \times 10^{-25}\ {\rm cm}^2$ is the Thomson cross section.
}. In this case it is evident that $L_p$ for single zone models we consider in this work violate $L_{\rm Edd}$ which is reasonable given that the absolute jet luminosity for blazars can exceed the accretion luminosity (generally a fraction of the Eddington luminosity)~\cite{Ghisellini:2014pwa}. The CR protons are isotropically accelerated within a blob, and the blob itself moves in the jet of TXS 0506+056 with speed $\beta_{B}$ with respect the central black hole. The corresponding Lorentz boost factor is given by $\Gamma_B \equiv\left(1-\beta_B^2\right)^{-1 / 2}$. For TXS 0506+056 we assume $\Gamma_B \approx 20$.


We follow the model "LMBB2b" from~\cite{Keivani:2018rnh} to simultaneously explain the neutrino and EM observations, where the CR proton power-law index $\alpha_p = 2$. The Lorentz factor $\Gamma_B \approx 20$ which results in an isotropic equivalent CR proton luminosity (in the observer frame) $L_p = \Gamma_B^4 L_p^{\prime} = 8.6 \times 10^{49}\ \rm erg\ s^{-1}$. Based on the model, the neutrino emission peak occurs at 10 PeV which corresponds to $\gamma_{p,\rm max}^\prime = 1.6 \times 10^7$, where primed coordinates refer to boosts in the frame of the blob. As for the low energy cut we assume $\gamma_{p,\rm min}^\prime = 1$. The comoving blob radius is given by $R_B^\prime \approx \Gamma_B c t_{\rm var}/(1+z) \sim 300 R_s$, where the Schwarzschild radius $R_s = 2GM_{\rm BH}/c^2$, $t_{\rm var}$ is the variability timescale of the source. Thus, the acceleration radius $R_{\rm acc} \sim \Gamma_B R_B/{\rm tan}\ \theta_j$, where $\theta_j \sim 1/\Gamma_B$ which gives $R_{\rm acc} \sim \Gamma_B^3 c t_{\rm var}/(1+z) \approx 1.8 \times 10^{19}$ cm ($\sim 6$ pc). Based on the X-ray and gamma-ray observations the variability timescale is $t_{\rm var} \lesssim 10^5$ s. The CR proton spectra is shown in Fig.~\ref{fig:ngc1068_crlum} (\textit{left}) as a solid blue line.
%
%
\subsection{NGC 1068}
\label{subsec:ngccrspec}
The IceCube Collaboration also reported an excess of $79^{+22}_{-20}$ neutrinos at $4.2\sigma$ significance associated with NGC 1068 - a Seyfert Type-II galaxy at a distance of $\sim 10$ Mpc~\cite{IceCube:2022der}. The galaxy consists of a central black hole with mass $M_{\rm BH} = 6\times 10^6 M_\odot$ along with an accretion disk. The production of high-energy neutrinos and consequently gamma-rays is mostly attributed to hadronic processes - hadronuclear ($pp$) and photomeson ($p\gamma$) interactions, resulting in the production of mesons. The observed isotropic equivalent neutrino luminosity is $L_\nu \simeq 1.4 \times 10^{42}\ \rm erg\ s^{-1}$ for neutrino energy $E_\nu$ between 1.5 TeV to 15 TeV. The isotropic equivalent gamma-ray luminosity between 100 MeV and 100 GeV is $7.7 \times 10^{40}\ \rm erg\ s^{-1}$.
Such high-energy emissions naturally require sites of efficient particle acceleration. The various mechanisms for such particle acceleration include magnetically-powered corona models via turbulence and magnetic reconnection, shocks, and reconnection-driven flows~\cite{Murase:2019vdl,Murase:2022dog,Fiorillo:2023dts,Das:2024vug,Fiorillo:2024akm}.

The IceCube neutrino observations from~\cite{IceCube:2022der} (see Fig. 4 there) are converted into corresponding CR proton energies by assuming $T_p \approx 20 T_\nu$ as before. The differential CR proton spectrum is approximately related to the neutrino spectrum by Eq.~\eqref{eq:diffptonu}. For a conservative estimate we take $K=1$ corresponding to the $p\gamma$ scenario, where the system is calorimetric such that $f_{p\gamma} \sim 1$~\cite{Murase:2019vdl,Murase:2022dog,Das:2024vug}. This gives the best fit range (blue shared region) in the right plot of Fig.~\ref{fig:ngc1068_crlum}, while the light blue band between the dashed blue lines cover the power law CR proton spectra that are consistent with the neutrino observations at $95\%$ C.L. The best-fit spectral index for the CR protons is given by $\alpha_p = 3.2$ (see Eq.~\ref{eq:crlum}). Given this index, the acceleration radius ($R_{\rm acc}$) is $\sim 30\ R_s$. This is consistent with the requirements of the minimum CR luminosity~\cite{Murase:2022dog,Das:2024vug,Murase:2019vdl} in case of both the $pp$ and $p\gamma$ scenarios. Note that the high-energy neutrino production efficiency is inversely proportional to the acceleration radius $R_{\rm acc}$, thus a smaller radius leads to more efficient neutrino production with a smaller value of $L_p$.
We explore two cases corresponding to the CR spectra for NGC 1068 which are distinguished by the normalization of the CR luminosity ($L_p$), the minimum ($T_p^{\rm min}$), and maximum ($T_p^{\rm max}$) CR proton kinetic energy.
\begin{itemize}
\item Case 1 (C1): X-ray observations for the total coronal luminosity above $2$ keV is given by $L_X = 7^{+7}_{-4} \times 10^{43} {\rm erg\ s}^{-1}$. We adopt this as a representative case where we normalize the CR luminosity (see Eq.~\ref{eq:crlum}) to $L_X$. Note that this is still conservative given that the contributions from the optically thick and geometrically thin disk is such that the bolometric luminosity is $L_{\rm bol} \simeq 4.8 \times 10^{44} {\rm erg\ s}^{-1}$. The power law index ($\alpha_p$) is consistent with the IceCube best fit. We choose $T_p^{\min} = 10$ TeV and $T_p^{\rm max} = 30$ TeV. Given that the IceCube neutrino flux has a lower limit of $E_\nu \sim 1$ TeV (implying $T_p \sim 20$ TeV), our chosen value of $T_p^{\rm min}$ is reasonable. On the other hand, the choice of $T_p^{\rm max}$ is conservative since a larger value would imply more neutrinos with higher energies leading to luminosities greater than $L_X$. This case is shown as the solid red line in Fig.~\ref{fig:ngc1068_crlum}.
\item Case 2a (C2a): For this case, we choose $T_p^{\rm max} \sim 300$ TeV according to the upper limit from the observed IceCube neutrino flux. The lower limit is chosen such that the normalization $L_p \approx L_{\rm edd}$, where the Eddington luminosity is given by $L_{\rm edd} \simeq 7.7 \times 10^{44} {\rm erg\ s}^{-1}$ and shown as the black dotted line in Fig.~\ref{fig:ngc1068_crlum}. This results in $T_p^{\rm min} \sim 100$ GeV. The spectral index for the CR protons match the IceCube best fit value of $3.2$.  This can in-principle mimic the $pp$ scenario for neutrino production. We show this case using a purple dashed line in Fig.~\ref{fig:ngc1068_crlum}. Note that choosing a harder spectrum (resulting from magnetic reconnection and/or stochastic acceleration) like $\alpha_p = 2$, would in turn lead to a larger value of $T_p^{\rm min}$ since $L_p \lesssim L_{\rm edd}$. 

\item Case 2b (C2b): We also consider a small variation of the above C2a case where instead of extrapolating the IceCube best fit to lower energies, we just use the IceCube results for the best fit (solid blue line) between $E_p^{\rm min} \sim 30$ TeV and $T_p^{\rm max} \sim 300$ TeV (same as C2a). The normalization for this case is $L_p \simeq 1.2 \times 10^{42} {\rm erg\ s}^{-1}$. This serves as the most conservative scenario.
\end{itemize}
We discuss the results corresponding to C1 in the main text, while the cases C2a and C2b are discussed in Appendix \ref{sec:NGC_Cases}. The two cases described above complement each other and attempts to fold in the model uncertainties associated with the astrophysical modeling of NGC 1068, including, sites and mechanisms of particle acceleration, neutrino production processes and the resulting spectra. The two cases thus help explore the non-trivial effects on the constraints as will be evident in the following sections.\\
\subsection{Spectral Modeling}
The spectral distribution of CRs on Earth's frame corresponds to a power-law function with energy. We follow the formalism from \cite{Wang:2021jic, Granelli:2022ysi}, such that the CR flux reads
\begin{equation}
\frac{d \Gamma_p}{d T_p d \Omega}=\frac{1}{4 \pi} c_p\left(1+\frac{T_p}{m_p}\right)^{-\alpha_p} \frac{\beta_p\left(1-\beta_p \beta_B \mu\right)^{-\alpha_p} \Gamma_B^{-\alpha_p}}{\sqrt{\left(1-\beta_p \beta_B \mu\right)^2-\left(1-\beta_p^2\right)\left(1-\beta_B^2\right)}} ,
\label{eq-AGN-Flux}
\end{equation}
If we want $\frac{d \Gamma_{p}}{dT_{p}}$ as used in Eq. \ref{eq:crlum}, then we integrate over solid angle $\Omega$. In the above equation the subscript $p$ is for protons with mass $m_p \simeq 0.938$ GeV, $\alpha_p$ is the spectral index of the CR proton flux, $T_p$ and $\beta_p=$ $\left[1-m_p^2 /\left(T_p+m_p\right)^2\right]^{1 / 2}$ are respectively the kinetic energy and velocity of the particle, $\mu$ is the cosine of the angle between the particle's direction of motion and the jet, and $c_i$ is a normalization constant proportional to the source luminosity $L_i$, via
\begin{equation}
L_p=\int d \Omega \int d T_p\left(T_p+m_p\right) \frac{d \Gamma_p}{d T_{p} d \Omega}=c_i m_p^2 \Gamma_B^2 \int_{\gamma_{\min , p}^{\prime}}^{\gamma_{\max , p}^{\prime}} d \gamma_p^{\prime}\left(\gamma_p^{\prime}\right)^{1-\alpha_i},
\end{equation}
thus
\begin{equation}
c_p=\frac{L_p}{m_p^2 \Gamma_B^2} \times \begin{cases}\left(2-\alpha_p\right) /\left[\left(\gamma_{\max , p}^{\prime}\right)^{2-\alpha_p}-\left(\gamma_{\min , p}^{\prime}\right)^{2-\alpha_p}\right] & \text { if } \alpha_p \neq 2 ; \\ 1 / \log \left(\gamma_{\max , p}^{\prime} / \gamma_{\min , p}^{\prime}\right) & \text { if } \alpha_p=2 ,\end{cases}
\end{equation}
with $\gamma^{\prime}$ the particle Lorentz factor in the frame of the blob. In Table \ref{fig:table1}, we show the relevant parameters considered in this paper for TXS 0506+056 and NGC 1068. Since NGC 1068 does not show evidence for a jet, we can set $\Gamma_{B} = 1 , \,(\beta_{B} = 0)$ and use the same equations. We use these parameters to show the results in Fig. \ref{fig:ngc1068_crlum}. We note that at high energies, our spectrum follows a power-law, while for low energies at TXS 0506+056, the flux becomes suppressed. This is due to the jet boost factor: even protons which are low energy in the frame of the blob are boosted to relativistic energies. This effect becomes strong at $T_{p} \lesssim \Gamma_{B} m_{p}.$ 
\section{Dark matter distribution in Active Galactic Nuclei \label{sec:dm_agn}}

Now we turn towards discussing the density profile of DM particles at the source and the corresponding column density that CRs traverse before escaping the source, while boosting DM on its way. It has long been known that adiabatically-growing black holes are expected to form a spike of DM particles in their vicinity \cite{1972GReGr...3...63P,Quinlan:1994ed, Gondolo_1999}. An initial DM profile of the form $\rho (r) = \rho_0 (r/r_0)^{-\gamma}$ evolves into
\begin{align}
	\rho_{\rm sp}(r) = \rho_{R} \, g_{\gamma}(r)\, \bigg(\frac{R_{\rm sp}}{r}\bigg)^{\gamma_{\rm sp}}\;,
\end{align}
where $R_{\rm sp}=\alpha_{\gamma}r_0(M_{\rm BH}/(\rho_{0}r_{0}^{3}))^{\frac{1}{3-\gamma}}$ is the size of the spike, with $\alpha_\gamma\simeq 0.293\gamma^{4/9}$ for $\gamma \ll 1$, and numerical values for other values of $\gamma$ provided in \cite{Gondolo_1999}. The steepness of the profile is parameterized by\footnote{The steepness of the spike is robustly predicted from dimensional arguments. Let's assume a model consisting entirely of circular orbits. Further, let's assume that the initial density cusp is $\rho \sim r^{-\gamma}$ before the addition of the black hole and $\rho \sim r^{-\gamma_{\mathrm{sp}}}$ after the black hole has accreted most of its mass. Conservation of mass implies that $\rho_i r_i^2 d r_i=\rho_f r_f^2 d r_f  \Longrightarrow r_i^{3-\gamma} \sim r_f^{3-\gamma_{\mathrm{sp}}}$. Conservation of angular momentum implies that $r_i M_i(r)=r_f M_f(r) \simeq r_f M_{\rm BH} \Longrightarrow  r_i^{4-\gamma} \sim r_f $. Combining these two results, one gets $\gamma_{\mathrm{sp}}=\frac{9-2 \gamma}{4-\gamma}$.} $\gamma_{\rm sp}=\frac{9-2\gamma}{4-\gamma}$, and $r_0$ denotes the scale radius of the host galaxy. Furthermore, $g_{\gamma}(r)$ is a function which  can be approximated for $0<\gamma <2 $ by  $g_{\gamma}(r) \simeq (1-\frac{2R_{\rm S}}{r})$, with $R_{\rm S}$ being the Schwarzschild radius, while $\rho_R=\rho_{0}\, (R_{\rm sp}/r_0)^{-\gamma}$.

We consider that the initial DM distribution follows a NFW profile \cite{Navarro:1996gj,Navarro:1995iw} with $\gamma=1$, resulting in a spike with $\gamma_{\rm sp}=7/3$ and $\alpha_{\gamma}= 0.122$ \cite{Gondolo_1999} (although we note that if the dark matter distribution near the galactic center is a core as suggested in \cite{Salucci:2018hqu}, this would only mildly change the spike index to 2.25). The masses of the central SMBHs of the two AGN considered in this work are given in Table \ref{fig:table1}. The normalization factor ($\rho_0$) can be obtained from the uncertainty in the black hole mass \cite{Gorchtein:2010xa,Lacroix_2017}, in such a way that the profile is compatible with both the total mass of the galaxy and the mass enclosed within the radius of influence of the $\mathrm{BH}$, of order $10^5 R_{\mathrm{S}}$. We follow in this paper the criteria from  \cite{Gorchtein:2010xa}. The DM spike mass within the region that is relevant for the determination of the BH mass, typically $R_0=10^5 R_{\mathrm{S}}$, must be smaller than the uncertainty on the $\mathrm{BH}$ mass $\Delta M_{\mathrm{BH}}$. For NGC 1068, we adopt $M^{\rm NGC}_{\rm BH}=(1-2) \times 10^7 M_{\odot}$ \cite{Woo:2002un}, and for TXS 0506+056, we take $M^{\rm TXS}_{\rm BH}=(3-10) \times 10^8 M_{\odot}$ \cite{Padovani:2019xcv}. The normalization constant $\rho_0$ is thus obtained by solving the following equation

\begin{equation}
\int_{4 R_{\mathrm{S}}}^{R_0} 4 \pi r^2 \rho_{\rm sp}(r) \mathrm{d} r=\Delta M_{\mathrm{BH}}.
\end{equation}

Factorizing $\rho_0$ in this expression we find a general expression

\begin{equation}
\rho_0=\left[\frac{\Delta M_{\mathrm{BH}}}{4 \pi\left(\alpha_{\gamma} r_0\left(M_{\mathrm{BH}} / r_0^3\right)^{\frac{1}{3-\gamma}}\right)^{\gamma_{\mathrm{sp}}-\gamma} r_0^{\gamma} \int_{4 R_S}^{R_0} g_\gamma(r) r^{2-\gamma_{\mathrm{sp}}} d r}\right]^{4-\gamma}
\end{equation}
where we have used the fact that $\frac{\gamma -3}{\gamma_{sp} -3} = 4 -\gamma$ to simplify the overall exponent. This expression still contains a non-trivial integral over $r$. It has previously been assumed in the literature that $g_{\gamma}(r) \sim 1$ (only true for $r \gg R_S$), and considered that the mass is dominated by the contribution from $r \gg R_{\mathrm{S}}$, i.e., typically $r>R_{\min }=\mathcal{O}\left(100 R_{\mathrm{S}}\right)$ \cite{Gorchtein:2010xa,Lacroix_2015}. One can then obtain a simpler expression

\begin{equation}
\rho_0 \simeq\left[\frac{\left(3-\gamma_{\mathrm{sp}}\right) \Delta M_{\mathrm{BH}}}{4 \pi \left(\alpha_{\gamma} r_0\left(M_{\mathrm{BH}} / r_0^3\right)^{\frac{1}{3-\gamma}}\right)^{ \gamma_{\mathrm{sp}}-\gamma} r_0^\gamma\left(R_0^{3-\gamma_{\mathrm{sp}}}-R_{\min }^{3-\gamma_{\mathrm{sp}}}\right)}\right]^{4-\gamma} .
\end{equation}

This criteria, applied to $\gamma=1$, yields masses of the DM halo below as expected from universal relations between supermassive black hole (SMBH)-galaxy bulge masses \cite{DiMatteo:2003zx,Ferrarese:2002ct}: 
\begin{equation}
\frac{M_{\mathrm{BH}}}{10^8 \mathrm{M}_{\odot}} \sim 0.7\left(\frac{M_{\mathrm{\chi}}}{10^{12} \mathrm{M}_{\odot}}\right)^{4 / 3}.
\end{equation}
If the DM particles self-annihilate, the maximal DM density in the inner regions of the spike is saturated to $\rho_{\text {sat}}= m_{\chi} /(\langle\sigma v \rangle t_{\mathrm{BH}})$, where $\langle \sigma v \rangle$ is the velocity averaged DM self-annihilation cross section, and $t_{\rm BH}$ is the SMBH age. We assume in this work that (co-)annihilations are not efficient in depleting the DM spike and $\rho_{\rm sat} \gg \rho_{\rm sp}$.
Furthermore, the DM spike extends to a maximal radius $R_{\rm sp}$, beyond which the DM distribution follows the initial NFW profile. The DM density profile therefore reads \cite{Gondolo_1999, Lacroix_2015, Lacroix_2017}
\begin{align}
        \rho(r)= \begin{cases} 
		0 & r\leq 4R_{\rm S}, \\
		\frac{\rho_{\rm sp}(r)\rho_{\rm sat}}{\rho_{\rm sp}(r)+\rho_{\rm sat}} & 4R_{\rm S}\leq r\leq R_{\rm sp}, \\
		\rho_{0}\left(\frac{r}{r_0}\right)^{-\gamma} \left(1+\frac{r}{r_0}\right)^{-(3-\gamma)} & r\geq R_{\rm sp}
        \end{cases}\,.
	\label{eq:spike_profile}
\end{align}
From the DM distribution we can find the column density of DM particles at these sources analytically. For the accelerating region of CRs, $R_{\rm acc}$, we take $R^{\rm NGC}_{\rm acc}=30 R_S^{\rm NGC}$ \cite{Murase:2022dog} and $R^{\rm TXS}_{\rm acc}=1.2\times10^5 R_S^{\rm TXS}$ \cite{Keivani:2018rnh} (see Table~\ref{fig:table1}), consistent with inferred values from multi-messenger observations from these sources , and with our CR modeling descriptions in Sec.~\ref{sec:AGN}. The exact column density in the spike reads 
\begin{equation}
\Sigma_{\chi, \mathrm{spike}}=\rho_{\mathrm{sp}}\left(R_{\mathrm{acc}}\right) \frac{R_{\mathrm{acc}}}{g_\gamma\left(R_{\mathrm{acc}}\right)} \int_1^{R_{\mathrm{sp}} / R_{\mathrm{acc}}} d y g_\gamma\left(R_{\mathrm{acc}} y\right) y^{-\gamma_{\mathrm{sp}}},
\end{equation}
with $y = r / R_{\mathrm{acc}}$. Dropping the $g_{\gamma}(r)$-dependence, one gets the approximation \cite{Ferrer:2022kei}
\begin{equation}
\Sigma_{\chi, \mathrm{spike}} \simeq \int_{R_{\rm acc}}^{R_{\rm sp}} dr \rho_{\rm sp}(R_{\rm acc})\left(\frac{r}{R_{\rm acc}}\right)^{-\gamma_{\rm sp}}\simeq \frac{\rho_{\rm sp}(R_{\rm acc}) R_{\rm acc}}{(\gamma_{\rm sp}-1)} \left[1-\left(\frac{R_{\rm sp}}{R_{\rm acc}}\right)^{1-\gamma_{\rm sp}}\right]\;.
\label{eq:SigmaSpike}
\end{equation}
Further, the contribution to $\Sigma_{\chi}$ from the passage through the halo of the host galaxy is
\begin{equation}
\Sigma_{\chi, \mathrm{host}}=\rho_0 r_0\left[\ln \left(1+\frac{r_0}{R_{\mathrm{sp}}}\right)-\frac{1}{1+R_{\mathrm{sp}} / r_0}\right] \simeq \rho_0 r_0 \Big[\log\left(\frac{r_0}{R_{\rm sp}}\right)-1\Big]\;,
\end{equation}
where the last approximation assumes $r_0\gg R_{\rm sp}$. Under this prescription, we find that the column density of DM particles in the spike (with $\gamma=1$) of TXS 0506+056 is given by $\Sigma^{\rm TXS}_{\chi}=1.9 \times 10^{28}$ GeV cm$^{-2}$, and for NGC 1068, we find $\Sigma^{\rm NGC}_{\chi}=1.7 \times 10^{33}$ GeV cm$^{-2}$. We will use these fiducial values along the paper. The corresponding boosted fluxes would change linearly with the column density of DM traversed at the source.
\section{Cosmic-ray boosted dark matter flux\label{sec:Boostedfluxes}}
We consider a coupling between the SM and the dark sector of the form
\begin{equation}
\mathcal{L} \supset  -\frac{1}{4} F_{\mu \nu}^{\prime} F^{\prime \mu \nu} +\frac{m_{Z^{\prime}}^2}{2} Z_\mu^{\prime} Z^{\prime \mu}+ig_{\chi}Z^{\prime \mu} \bar{\chi}_1 \gamma_\mu \chi_2 + (h.c) + i \sum_{f} g_{f} Z^{\prime \mu} \bar{f} \gamma_{\mu} f,
\end{equation}
with $F^{\prime \mu \nu}=\partial^\mu Z^{\prime \nu}-\partial^\nu Z^{\prime \mu}$ and the sum in the last term being over SM fermions. In what follows we will examine our constraints on two classes of models. In the first case, we adopt the assumption that the $Z'$ couples equally to protons, neutrons, and electrons,  $g_{e} = g_{p} = g_{n} =  3g_{q}$, and set other SM couplings to zero. In the second case, we will examine the dark photon scenario in which $g_{e} = -g_{p}$, with all other SM couplings set to zero. Given that the neutron coupling plays a minor role in the phenomenology we consider, both model predictions can be done simultaneously. The dark sector contains a vector mediator $Z^{\prime}$, a stable Dirac fermion DM particle $\chi_{1}$ of mass $m_{\chi}$, and an excited Dirac fermion $\chi_2$ of mass $m_{\chi} + \delta$. The DM ($\chi_1$) scattering cross section with CRs $i=p,e$ ($\chi_1+i \rightarrow \chi_2 +i $) is given by 
\begin{equation}
    \frac{d \sigma_i}{dT_{\chi_{2}}} = \frac{\sigma_{\chi-i} }{4}\frac{m_{Z^{\prime}}^4}{(m_{Z^{\prime}}^2 + q^2)^2} \frac{m_{\chi} \bigg[ \big(s - (m_{\chi}^2 + m_{i}^2 + \delta m_{\chi}) \big)^2 + m_{\chi} T_{\chi_{2}} (q^2 - 2s) \bigg] }{2 \mu_{\chi-i}^{2} \lambda(m_{\chi}^2,m_{i}^2,s)} F_{i}^2(q^2),
    \label{eq:Vec_Diff_Sigma}
\end{equation}
where $T_{\chi_{2}}$ is the upscattered DM kinetic energy, the momentum transfer is $q^2=2 m_{\chi} T_{\chi_{2}}-\delta^2$, the  Mandelstam variable $\mathrm{s}=m_i^2+m_{\chi}^2+2\left(m_i+T_{i}\right) m_{\chi}$, $\lambda(a, b, c)=a^2+b^2+c^2-2 a b-2 a c-2 b c$. The non-relativistic DM-proton and DM-electron scattering cross section is given by
\begin{equation}
    \sigma_{\chi-i} = 4\frac{g_{i}^2 g_{\chi}^2 \mu^2_{\chi-i}}{\pi m_{Z^{\prime}}^4},
    \label{eq:sigma_NR}
\end{equation}
where $g_{i}$ denotes the gauge coupling of the mediator $Z^{\prime}$ to protons or electrons, and $g_{\chi}$ denotes the coupling of the mediator to the DM. The reduced mass of the DM-proton/electron system is $\mu_{\chi-i}$. In models with a vector portal between the DM and the SM sectors, the gauge coupling $g_i$ is typically proportional to the kinetic mixing $\epsilon$ between the vector boson and the SM photon \cite{Holdom:1985ag}. The non-relativistic cross section from Eq. \ref{eq:sigma_NR} can then be written as
\begin{equation}
\sigma_{\chi-i}=\frac{64 \pi \mu_{\chi-i}^2 \alpha \alpha_D \varepsilon^2}{m_{Z^{\prime}}^4} \label{eq-sigma-dark-photon},
\end{equation}
where $\alpha_D=g_{\chi}^2/4\pi$ is the dark electromagnetic structure constant. The from factor $F_i\left(q^2\right)$ can be for either the electron, quark, or proton. For electrons and quarks, it equals 1, while for protons it reads
\begin{equation}
F_p\left(q^2\right)=\frac{1}{\left(1+q^2 / \Lambda^2\right)^2}
\end{equation}
where $\Lambda \simeq 770$ MeV \cite{ANGELI2004185}. We note that while this form factor is simple, any corrections will enter at $\mathcal{O}(q^2/m_{p}^2)$ or higher \cite{DeMarchi:2025uoo}. The form factor above already suppresses scattering at these momentum transfers, so any corrections will not change the overall phenomenology, at most altering behavior by $\mathcal{O}(1)$ in a region where our constraints are weak. For now, we will consider only elastic proton scattering, although we will comment on DIS later on (resonances are ignored entirely, which is a mildly conservative approximation, since either elastic scattering or DIS dominate over most energy ranges considered in this work).

We know that the initial scattering angle is uniquely determined by the proton and $\chi_{2}$ kinetic energies
\begin{equation}
    \cos\theta(T_p,T_{\chi_2}) = \frac{-\delta^2 + 2 m_{p} (\delta + T_{\chi_2}) + 2 m_{\chi} T_{\chi_2} + 2\delta T_{p} + 2 T_{p} T_{\chi_2}}{2 \sqrt{T_{p} (2 m_{p} + T_{p})} \sqrt{T_{\chi_2} (2 \delta + 2 m_{\chi} + T_{\chi_2})}}.
    \label{eq:Scattering_Angle}
\end{equation}

\subsection{Decays}

We now introduce DM decays, and assume the decay length is small compared to intergalactic distances, so all interactions can be assumed to happen at the AGN. We will consider the following two-body and three-body decay processes
\begin{equation}
\chi_{2} \rightarrow
\begin{cases}
     \chi_{1} + Z^{\prime} \, \, \mathrm{if} \, \, \delta > m_{Z^{\prime}} \\
    \chi_{1} + e + \Bar{e} \, \, \mathrm{if} \, \, \delta < m_{Z^{\prime}} \, \& \, \delta > 2 m_{e} \\
    \chi_{1} + \gamma \, \, \mathrm{if} \, \, \delta < m_{Z^{\prime}} \,\& \, \delta < 2 m_{e}\,
\end{cases}\,.
\end{equation}
In order to account for such decay processes, let us consider a proton which is propagating at angle $\theta_{p}$ relative to the jet (where the jet iself is pointed at Earth). The proton upscatters a $\chi_{2}$ particle at scattering angle $\theta_{s}$ and azimuthal angle $\phi_{s}$. The $\chi_{2}$ then decays into a $\chi_{1}$ which propagates at angle $\theta_{d}$ relative to the parent particle. For the $\chi_{1}$ particle to reach Earth, the geometry requires
\begin{equation}
    \cos\theta_{p} = \cos\theta_{s} \cos\theta_{d} + \sin\theta_{s} \sin\theta_{d} \cos\phi_{s}.
\end{equation}
As indicated in Eq. \ref{eq:Scattering_Angle} the scattering angle $\theta_{s}$ is uniquely determined by $T_{p}$ and $T_{\chi_2}$. When considering decays, it is helpful to work in the rest frame of the $\chi_{2}$ particle, so we can define the angle, energy, and momentum of $\chi_{1}$ in this frame to be $\theta_{r}$, $E_{\chi_{1},r}$, $p_{r}$ respectively. Letting $\gamma = 1 + T_{\chi_{2}}/(m_{\chi} + \delta)$ and $\beta = \sqrt{1 - 1/\gamma^2}$, we can find quantities in the lab frame as
\begin{equation}
    E_{\chi_{1}} =  \gamma E_{\chi_{1},r} + \gamma \beta p_{r} \cos \theta_{r},
\end{equation}
and
\begin{equation}
    \cos\theta_{d} = \frac{\gamma \beta E_{\chi 1,r} + \gamma p_{r} \cos \theta_{r}}{\sqrt{E_{\chi_{1}}^2 - m_{\chi}^2}}.
\end{equation}
Inverting these expressions we find
\begin{equation}
    \cos\theta_{r} = \frac{E_{\chi_{1}} - \gamma E_{\chi_{1}, r}}{\gamma \beta p_{r}}.
\end{equation}
By considering the double differential decay rate of $\chi_{2}$, $\frac{d^2\Gamma_{\chi}}{d\cos\theta_{r} dE_{\chi_{1},r}}$, with total decay rate $\Gamma_{\chi}$, we may now compute the flux as
\begin{equation}
    \frac{d \Phi_{\chi_{1}}}{d E_{\chi_{1}}} = \frac{\Sigma_{\chi_{1}}}{2 \pi m_{\chi} d^2} \int d\cos\theta_{r}d\phi_{s} dT_{\chi_{2}}  dT_{p} dE_{\chi_{1}, r} \frac{d \Gamma_{p}}{d T_{p} d \Omega} (\theta_{p}) \frac{d\sigma}{dT_{\chi_{2}}} \frac{1}{\Gamma_{\chi}} \frac{d^2 \Gamma_{\chi}}{d \cos\theta_{r} dE_{\chi_{1},r}} \delta(E_{\chi_{1}} - E_{\chi_{1}}(T_{\chi_{2}},\theta_{r},E_{\chi_{1},r})).
\end{equation}
For all of decays considered, we assume that they are isotropic in the rest frame. We can then integrate out $\cos\theta_{r}$
\begin{equation}
\begin{split}
    \frac{d \Phi_{\chi_{1}}}{d E_{\chi_{1}}} \bigg|_{iso} =& \frac{\Sigma_{\chi_{1}}}{4 \pi m_{\chi} d^2} \int d\phi_{s} dT_{\chi_2}  dT_{p} dE_{\chi_{1},r} \frac{d \Gamma_{p}}{d T_{p} d \Omega} (\theta_{p}) \frac{d\sigma}{dT_{\chi_2}} \frac{1}{\Gamma_{\chi}} \frac{d \Gamma_{\chi_1}}{d E_{\chi_{1}, r}} \frac{1}{\gamma \beta p_{r}} \\
   & \Theta(E_{\chi_{1}} + \gamma \beta p_{r} - \gamma E_{\chi_{1},r}) \Theta(\gamma E_{\chi_{1},r} + \gamma \beta p_{r} - E_{\chi_{1}}).
   \label{eq-chi1-flux}
\end{split}
\end{equation}
As this assumption of isotropic decays continues for the rest of the text, we will drop the ``iso" subscript. For two-body decays, we find
\begin{equation}
    \frac{d \Gamma_{\chi}}{dE_{\chi_{1},r}} \bigg|_{\chi_{2} \rightarrow \chi_{1} + Z^{\prime}/\gamma} = \Gamma_{\chi} \delta \bigg( \frac{(m_{\chi} + \delta)^2 + m_{\chi}^2 - m_{Z^{\prime}/\gamma}^2}{2 (m_{\chi} + \delta) } - E_{\chi_{1},r} \bigg).
\end{equation}
In the case of three-body decays, we define the invariant mass square of the electron-positron pair as $m_{e \Bar{e}}^2 = (m_{\chi} + \delta)^2 + m_{\chi}^2 - 2 (m_{\chi} + \delta) E_{\chi_{1},r}$. The differential rate is then
\begin{equation}
    \frac{d \Gamma_{\chi}}{dE_{\chi_{1},r}} \bigg|_{\chi_{2} \rightarrow \chi_{1} + e + \Bar{e}} = -2(m_{\chi} + \delta)\frac{d \Gamma_{\chi}}{dm_{e \Bar{e}}^2},
\end{equation}
where
\begin{equation}
\frac{d \Gamma_{\chi}}{dm_{e \Bar{e}}^2} = \frac{g_{\chi}^2  g_{f}^2  \sqrt{m_{e \Bar{e}}^2 (m_{e \Bar{e}}^2 - 4 m_{e}^2)} (2 m_{e}^2 + m_{e \Bar{e}}^2) ( \delta^2 -m_{e \Bar{e}}^2)^{3/2} \sqrt{((\delta + 2 m_{\chi})^2 -m_{e \Bar{e}}^2) } ( 2 m_{e \Bar{e}}^2 + (\delta + 2 m_{\chi})^2)}{192 \pi^3 m_{e \Bar{e}}^2 (m_{e \Bar{e}}^2 - m_{Z^{\prime}}^2)^2 (\delta + m_{\chi})^3}.
\end{equation}
Following the described prescription, we show a set of simulated CR boosted DM fluxes for various combinations of free parameters in the left plot of Fig. \ref{fig:boosted_DM_fluxes}. The red (blue) lines indicate the flux from TXS 0506+056 (NGC 1068). We fix the DM and mediator mass and select gauge couplings to give a non-relativistic p-$\chi_{1}$ cross section (Eq. \ref{eq:sigma_NR}) of $10^{-30} \mathrm{cm^2}$. We find that the boosted DM fluxes peak at low DM energies, as expected given the falling power law spectra of CRs in AGN and the fact that scattering is suppressed when the momentum transfer is greater than the mediator mass. For the same choice of DM parameters, we notice that the boosted fluxes from NGC 1068 are larger than those from TXS 0506+056 at energies below $\sim 20$ GeV, while the order flips for higher energies. This behavior, along with the low-energy features on the boosted fluxes from TXS 0506+056, arise from its highly boosted jet.

\subsection{Deep-Inelastic Scattering (DIS)}

\begin{figure}[t!]
    \includegraphics[width = 0.45 \textwidth]{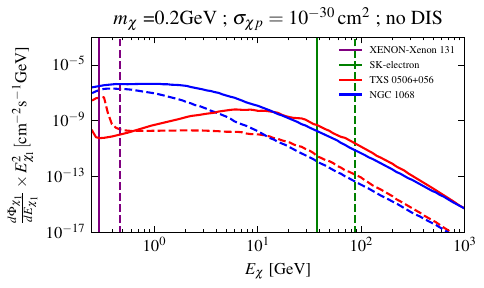}
    \includegraphics[width = 0.45 \textwidth]{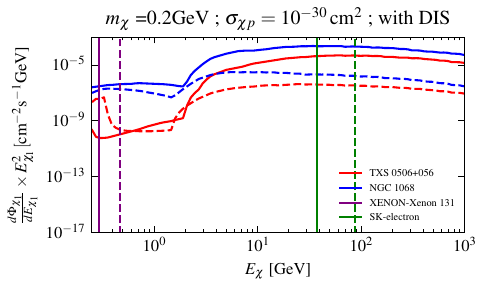}
    \caption{\justifying Boosted DM flux at Earth for a DM with $m_{\chi} = 0.2$ GeV and non-relativistic DM-proton cross section of $\sigma_{\chi p} = 10^{-30} \mathrm{cm^2}$. We show the results without (\textbf{left}) and with (\textbf{right}) DIS. Solid (dashed) lines indicate the fluxes when $m_{Z'} = 10 m_{\chi}$ and $\delta = 0.4 m_{\chi}$ ($m_{Z'} = 3 m_{\chi}$ and $\delta = 0.8 m_{\chi}$). We can see a difference in the fluxes from each source, as TXS 0506+056 has a jet, leading the fluxes to have a sharp peak at low energies. DIS leads to a substantial increase of the flux at high energies, as quarks are not subject to a form-factor supression. Vertical lines indicate the minimum energy necessary to scatter inelastically at XENONnT off Xenon-131 and Super-K off electrons, with solid and dashed lines indicating $\delta = 0.4 m_{\chi}$ and $\delta = 0.8 m_{\chi}$ respectively.\label{fig:boosted_DM_fluxes}}
\end{figure}

We can also consider the contribution from DIS. We will parametrize this in terms of the Bj{\"o}rken  parameter $x$. We must now use a double differential cross section $\frac{d^{2}\sigma}{dx dT_{\chi_{2}}}$. However, as we are working with high energy scatterings, we can use the approximation

\begin{equation}
    \frac{d^{2} \sigma}{dx dT_{\chi_{2}}} \xrightarrow[s \gg m_{p}^2, m_{\chi}^2]{} \sum_{i}f_{i}(x,q^2) \frac{d \sigma_{i}(x s)}{dT_{\chi_{2}}}
\end{equation}
where $f_{i}(x,q^{2})$ is the parton distribution obtained from \cite{Clark:2016jgm} and we sum over up, down, anti-up, and anti-down quarks. For $\frac{d \sigma_{i}}{dT}$, we use Eq \ref{eq:Vec_Diff_Sigma} with a form factor set to 1 and neglect the quark mass. The reason that we can take $s_{q} = x s$ where $s_{q}$ is the Mandelstam variable for the DM-quark interaction and $s$ is the variable for the overall DM-proton interaction is as follows. In the rest frame of the DM, the proton is highly boosted, and thus the four momentum of the proton $p_{p}^{\mu}$ and of the underlying quark $p_{q}^{\mu}$ are related by $p_{q}^{\mu} \simeq x p_{p}^{\mu} + \mathcal{O}(m_{p}/T_{p})$. Thus, when this kinetic energy is far higher than the masses of the particles involved,  we know $s_{q} = (p^{\mu}_{q} + p^{\mu}_{\chi})^2 \simeq x p_{p}^{\mu}p_{\chi,\mu} \simeq x s$. We can equivalently think of this as the quark having kinetic energy $T_{q} = \frac{1}{2 m_{\chi}} \big(s_{q}-(m_{q}+m_{\chi})^{2} \big)$ which allows us to compute the scattering angle via Eq. \ref{eq:Scattering_Angle}. In this approximation, we have

\begin{equation}
\begin{split}
    \frac{d \Phi_{\chi_{1}}}{d E_{\chi_{1}}} \bigg|_{\rm DIS} =& \frac{\Sigma_{\chi_{1}}}{4 \pi m_{\chi} d^2} \sum_{i} \int d\phi_{s} dT_{\chi_2}  dT_{p} dE_{\chi_{1},r} dx f_{i}(x,q^2) \frac{d \Gamma_{p}}{d T_{p} d \Omega} (\theta_{q}) \frac{d\sigma(xs)}{dT_{\chi_2}} \frac{1}{\Gamma_{\chi}} \frac{d \Gamma_{\chi_1}}{d E_{\chi_{1}, r}} \frac{1}{\gamma \beta p_{r}} \\
   & \Theta(E_{\chi_{1}} + \gamma \beta p_{r} - \gamma E_{\chi_{1},r}) \Theta(\gamma E_{\chi_{1},r} + \gamma \beta p_{r} - E_{\chi_{1}}),
   \label{eq-flux-DIS}
\end{split}
\end{equation}
where the sum is over up, down, anti-up, and anti-down quarks. We use Eq. \ref{eq:Vec_Diff_Sigma} with the form factor set to 1 for the cross section, find the scattering angles from the quark kinematics, and use parton distribution functions $f_{i}(x,q^2)$ from \cite{Hou:2019efy, Clark:2016jgm}. We consider that the coupling to quarks is 1/3 of the coupling to protons~\footnote{This is a conservative choice when extrapolating constraints to a dark photon model, as in that model up and down quarks have a coupling of 2/3 and -1/3 of the proton respectively.}. Also, we only consider DIS for $q > 1.295$ GeV, which is the lowest momentum transfer at which ManeParse includes parton distributions \cite{Clark:2016jgm}.
We also simulate the boosted DM fluxes at TXS 0506+056 and NGC 1068 including DIS as described. The results are shown in the right plot of Fig. \ref{fig:boosted_DM_fluxes}. We note that the flux from NGC 1068 stays above the TXS 0506+056 flux when DIS is considered, as there is no longer a form-factor suppression at high energies. Above a certain energy threshold when the momentum transfer can be above 1.295 GeV, the DIS contribution starts to contribute and the boosted DM fluxes are enhanced by orders of magnitude with respect to the case without DIS. The DM energy threshold for DIS to occur depends on the DM mass and mass splitting, thus in Fig.~\ref{fig:boosted_DM_fluxes} the fluxes present ``bumps" at different energies depending on the labeled assumptions. The effects of DIS become more important at Super-K than at experiments with lower energy thresholds such as XENONnT. The spectral features induced by DIS may also help to differentiate DM from other cosmogenic and neutrino-induced backgrounds.

%
\subsection{Attenuation from Earth \label{sec:Atten}}
The DM-SM cross sections considered here can be large enough that the Earth is no longer optically thin to the DM flux. In this case, we must now consider the effects of attenuation. To do so, knowing the location of both the AGN source and detector, we can find the optical depth $\tau$ as
\begin{equation}
\tau = \sum_{i} \sigma_{i} \int d\ell\ n_{i}(\ell)\,,
\end{equation}
where $i$ sums over the different elements, $\ell$ is a line-of-sight integral, and $n_{i}$ is the number density of each element \cite{PREM,CoreMantleComposition,  CrustComposition}. The cross section includes coherent scattering off nuclei, incoherent scattering on nucleons and electrons, and DIS (when considered at the source). We treat the detectors as 1 km underground in a spherically symmetric Earth. We note that the line of sight changes as the Earth rotates, and the cross section is an energy dependent quantity, so we cannot describe our problem with a single $\tau$.

To properly compute attenuation, we need to understand the interior of the Earth. For the density, we use the Preliminary Reference Earth Model \cite{PREM}. For the elemental compositions, we find the core/mantle composition from \cite{CoreMantleComposition} and the crust composition from \cite{CrustComposition}. The results are summarized in Table \ref{table:weights}.

\begin{table}[H]
\centering
$\begin{array}{ cccc}
\text{Element} 		& \text{Crust \%}  & \text{Mantle \%} & \text{Core \%} 	 \\
\hline


\text{O} & 46.6 &  44 & 0 \\ 
\text{Si} & 27.72 &  21 & 6 \\ 
\text{Al} & 8.13 &  2.35  & 0 \\ 
\text{Fe}  & 5.05 & 6.26 &  85.5 \\
\text{Ca} & 3.65 &  2.5 & 0 \\ 
\text{Na} & 2.75 & 0 & 0 \\
\text{K} & 2.58 & 0 & 0 \\
\text{Mg} & 2.08 & 22.8 & 0 \\
\text{S} & 0 & 0 & 1.9 \\
\text{Ni} & 0 & 0 & 5.2 \\
\hline
\hline
\text{Total} & 98.56 & 98.91 & 98.6 \\
 \hline
\end{array}$

\caption{Fractional weights for elements in each layer of the Earth
\label{table:weights}}
\end{table}

To set a ``ceiling" on our bounds, we consider a test energy

\begin{equation}
E_{\rm test} = \max \left[ m_{\chi}+ \delta +\Delta E, 1.1 \times \bigg( m_{\chi} + \frac{\delta^2 +2 \delta (m_{p} +m_{\chi})}{2 m_{p}} \bigg) \right],
\end{equation}
where $\Delta E = 1$ GeV (0.1 GeV) for Super-K (XENONnT) is chosen as a characteristic energy for detecting scattering signals, and the second value assures that upscattering is possible for DM-proton scattering. We set the ``ceiling" to be the minimum value of the couplings for which $\tau(E_{ \rm test}) \geq 1$ for all points during Earth's rotation. For such strong attenuation, the flux at our detectors would be too small and we would be unable to set constraints.

For less severe attenuation, we must update the flux to account for attenuation, which at the detector becomes
\begin{equation}
    \frac{d\Phi_{\chi_{1}}}{d E_{\chi_1}} \bigg |_{\rm det} = \langle \exp(-\tau(E_{\chi_1})) \rangle \frac{d \Phi_{\chi_{1}}}{dE_{\chi_1}} \bigg |_{\rm orig},
\end{equation}
where the brackets signify averaging over the Earth's rotation and the flux on the right-hand side is the flux without attenuation as obtained via Eq.~\ref{eq-chi1-flux} or Eq.~\ref{eq-flux-DIS} depending on if DIS is considered.
\section{Direct detection of cosmic-ray boosted DM at Super-K\label{sec:Rates}}
\begin{figure}[t!]
    \includegraphics[width = 0.45 \textwidth]{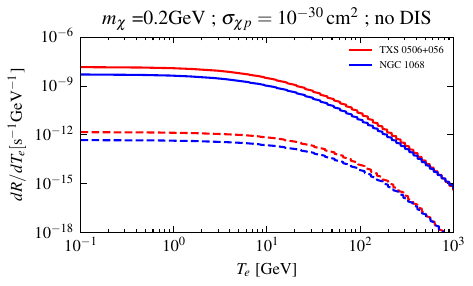}
    \includegraphics[width = 0.45 \textwidth]{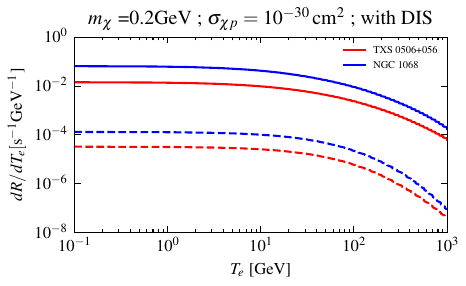}

    \caption{ \justifying Differential rate of scattering of DM with electrons at Super-K, with angular cuts on the events as described in the main text. Parameters are chosen to match Fig. \ref{fig:boosted_DM_fluxes}. As before, solid lines indicate $m_{Z'} = 10 m_{\chi}$ and $\delta = 0.4 m_{DM}$ while dashed lines indicate $m_{Z'} = 3 m_{\chi}$ and $\delta = 0.8 m_{\chi}$. \label{fig:Scat_Rate}}
\end{figure}

In the previous section, we have calculated the flux of $\chi_{1}$ particles reaching the Earth. The scattering rate at the detector protons, electrons or nuclei (denoted as $i=p,e,A$), is given by
\begin{equation}
    \frac{dR}{dT_{i}} = N_{i}  \int dE_{1} \frac{d \Phi_{\chi_1}}{dE_{\chi_1}} \frac{d \sigma_{1}(E_{1})}{dT_{i}}.
\label{eq:diff_recoil_rate}
\end{equation}
where $N_i$ is the number of target protons or electrons in the detector, and $T_i$ denotes their recoil energy of the proton or electron, and $E_{1}$ is the incoming energy of the DM particle. The scattering cross section for $\chi_{1}$ with a particle $i$ is 
\begin{equation}
    \frac{d \sigma_{1}(E_{1})}{dT_{i}} = \frac{g^2_{i} g^2_{\chi} m_{\chi} F_{i}(q^{2})}{8 \pi m_{i} m^4_{Z} (m^2_{\chi} - E^{2}_{1})} \Big( -4 E_{1}^2 m_{i} + 4 E_{1} m_{i} T_{i} + 2 \delta E_{1} (\delta + 2 m_{\chi})+ 2 m^2_{i} T_{i} + \delta^2 m_{i} - 2 m_{i} T^2_{i} + 2 m^2_{\chi} T_{i} -\delta^2 T_{i} \Big),
    \label{eq-dsigma1_dTSM}
\end{equation}
where as before the form factor is 1 for electrons and quarks. Blazar-boosted DM has been studied at Super-K previously \cite{Granelli:2022ysi}. In it, they study electron scattering in the 22.5 kton water Cherenkov detector, breaking their study into 3 bins based on recoil energy: Bin 1 [0.1 GeV, 1.33 GeV], Bin 2 [1.33 GeV, 20 GeV], Bin 3 [20 GeV, 1 TeV].  Furthermore, they make angular cuts of $24^{\circ}$, $7^{\circ}$, and $5^{\circ}$ respectively for the 3 bins. We show the electron scattering rates induced by CR boosted DM at Super-K applying the angular cuts in Fig \ref{fig:Scat_Rate}. We notice that the scattering rate is enhanced at low energies, where the boosted fluxes are larger.
The aforementioned angular cuts were chosen to contain most outgoing electrons from elastic scattering. As the kinematics for inelastic DM are different, we also consider the possible exclusions without the angular cuts, setting our 95\% confidence level where the number of DM-induced scatters $N_{\rm BSM} = 2 \sqrt{N_{\rm Bkg}}$ (with backgrounds obtained from \cite{SKBoostedElectron}). This allows us to set our limits at 20, 4.5, and 3 (150, 51, and 6) events with (without) angular cuts over 2500 days for Bins 1, 2, and 3 respectively. We assume that the detector has perfect efficiency, energy resolution, and angular resolution so that all events appear in the true angular and energy bins.
Constraints can also be placed through interactions with protons at the Super-K detector. Super-K has also analyzed proton data for CR boosted DM, looking for events with momentum above 1.2 GeV and below 3 GeV \cite{Super-Kamiokande:2022ncz}. In their sample, they predicted $111.7 \pm 10.6 (\mathrm{stat.}) \pm 30.7 (\mathrm{sys.})$ 
 (a relative $27 \%$ systematic uncertainty). As 126 events were observed, this means we can set a 90$\%$ confidence exclusion for any model producing 80 or more elastic proton-DM induced events. To do this, we consider just scattering off hydrogen atoms, ignoring oxygen.

We also consider DIS in the detector. Such events would produce multiple Cherenkov rings, so we can set a limit by using data in Super-K Run I-IV where 5755.9 multi-ring events are predicted compared to 5770 observed. Considering only statistical uncertainty, we would set $90\%$ confidence constraints at 114 DM-DIS events. We do not consider the energy distribution of DIS events but rather compute the entire rate as
 \begin{equation}
     R_{\rm DIS} =(N_p +N_{n}) \int dE_{1} \frac{d \Phi_{1}}{dE_{1}} \sigma_{1,\rm DIS}(s) 
 \end{equation}
where $N_{p}$ and $N_{n}$ are the number of protons and neutrons in the detector and
\begin{equation}
    \sigma_{1, \rm DIS}(s) = \sum_{i} \int dt dx f_{i}(x,-t) \frac{d \sigma_{1}(xs)}{dt}.
\end{equation}
where the sum is again over up, down, anti-up and anti-down quarks. We can relate $\frac{d\sigma(s)}{dt}$ to Eq. \ref{eq-dsigma1_dTSM}  via $\frac{d\sigma(s)}{dt} = \frac{1}{2 m_{i}} \frac{d\sigma}{dT_{i}}$ and using the substitutions $T_{i} \rightarrow -t/2m_{i}$ and $E_{1} \rightarrow \frac{1}{2 m_{i}} (s - m_{\chi}^2 - m_{i}^2)$ where $m_{i}$ is the quark mass. As before, we take the coupling of quarks to $Z^{\prime}$ to be 1/3 that of proton.

\begin{figure}
      \includegraphics[width=0.49\textwidth]{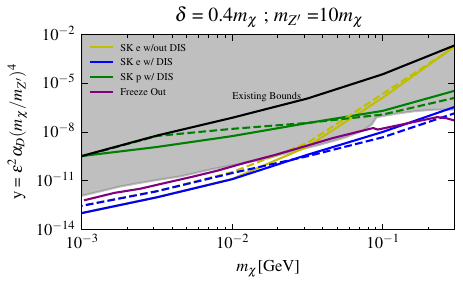}
      \includegraphics[width=0.49\textwidth]{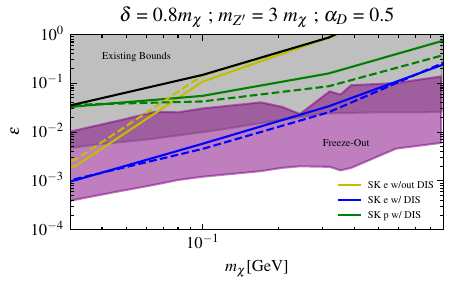}

    \caption{\justifying Excluded parameter space of light and inelastic DM sourced by TXS 0506+056 (solid lines) and NGC 1068 (dashed lines) from observations of electron and proton scattering at Super-K. We fix the mass splitting among the two DM states to be $\delta=0.4m_{\chi}$ (left panel) and $\delta=0.8m_{\chi}$ (right panel). We take a mediator with mass $m_{Z^{\prime}}=10m_{\chi}$ (left panel) and $m_{Z^{\prime}}=3m_{\chi}$ (right panel). The left panel shows constraints on the parameter $y$, DM and SM gauge coupling and masses of the DM and mediator. The right panel shows constraints solely on $\epsilon = g_{e}/e$ (see Eq. \ref{eq-sigma-dark-photon}). In both plots, we show the difference in the limits when considering elastic scatterings only and when including DIS, which enhances the sensitivity at large DM masses significantly. The black line indicates the ``ceiling", above which attenuation through the overburden significantly suppresses our signal. For comparison, we show existing bounds in this parameter space from colliders and beam dump experiments \cite{Garcia:2024uwf}, and bounds from CR cooling in AGN \cite{Gustafson:2024aom}. Furthermore, we show the region of parameter space favored by thermal freeze-out from \cite{Garcia:2024uwf}, accounting for a range of plausible values in the Majorana mass term chiral asymmetry (see also \cite{gonzález2021cosmology, Fitzpatrick:2021cij} for detailed calculations of the thermal relic abundance of inelastic DM).}
\label{fig:inelastic_light_limits}
\end{figure}

We will derive constraints on various combinations of the free BSM parameters for a proper assessment of the strength of this phenomenological probe of DM. In Fig.~\ref{fig:inelastic_light_limits}, we show constraints on light DM with mass splittings comparable to the DM mass, a scenario widely studied in the literature since it can be probed at collider and beam dump experiments, and where the thermal relic abundance has not been completely probed. We show constraints from electron scattering at Super-K, both with and without considering DIS at the AGN. We also include constraints from DIS scattering at Super-K when the same is also considered at the AGN. Constraints from elastic proton recoils are weaker than the attenuation ``ceiling" and are therefore not shown. As it can be appreciated in the figure, our bounds from CR boosted DM at NGC 1068 and TXS 0506+056 can improve over previous bounds from beam dump experiments and CR cooling in AGN, further testing the thermal freeze-out target for a wide range of inelastic DM masses. Concretely, Super-K is able to probe these models for DM masses $m_{\chi} \lesssim 0.1$ GeV ($\delta=0.4$m$_{\chi}$, $m_{Z^{\prime}}=10m_{\chi}$) and $m_{\chi} \lesssim 0.4$ GeV ($\delta=0.8_{\chi}$, $m_{Z^{\prime}}=3m_{\chi}$). It can also be appreciated in the plots that the inclusion of DIS enhances the sensitivity with respect to elastic scatterings at large DM masses. This is expected, as larger mass splittings require larger momentum transfers, where DIS has the higher cross section.

\begin{figure}[t!]
    \includegraphics[width = 0.49 \textwidth]{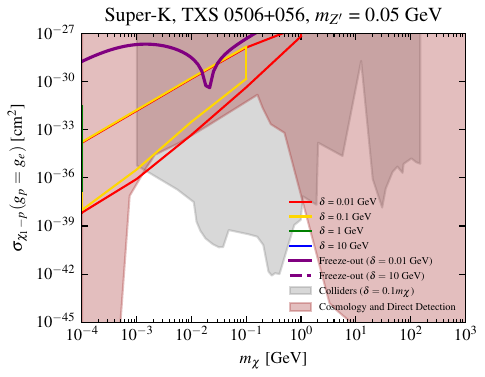}
    \includegraphics[width = 0.49 \textwidth]{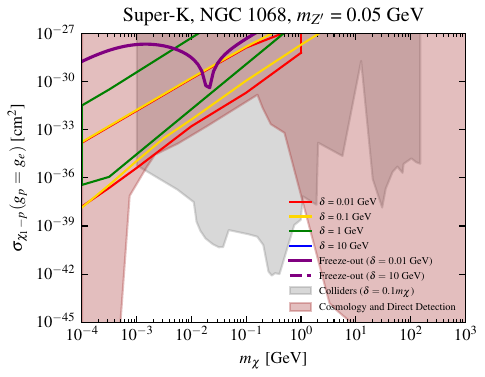}
    \includegraphics[width = 0.49 \textwidth]{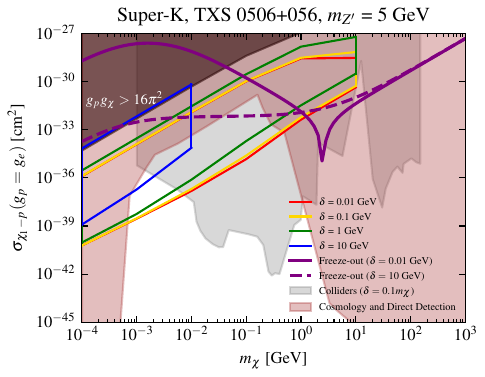}
    \includegraphics[width = 0.49 \textwidth]{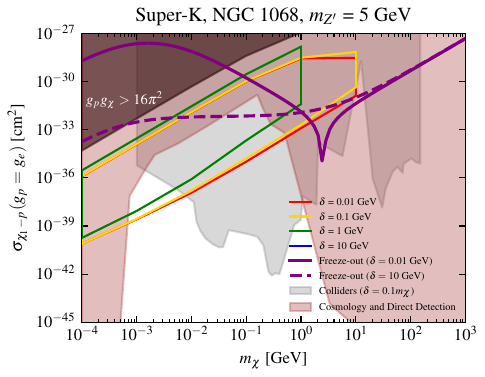}
    \includegraphics[width = 0.49 \textwidth]{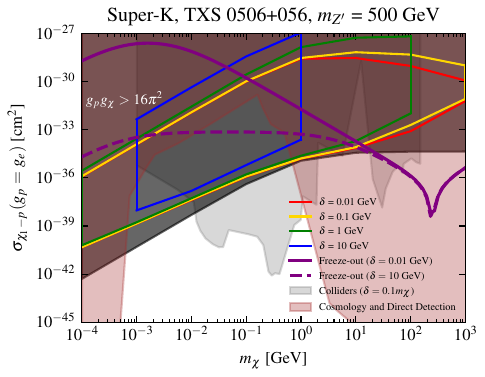}
    \includegraphics[width = 0.49 \textwidth]{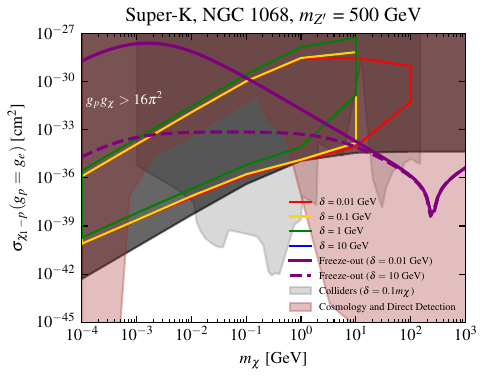}
    \caption{ \justifying Upper limits on the $\chi_1-p$ non-relativistic scattering cross sections ($\sigma_{\chi-p} = 4 g_{\chi}^2 g_{p}^2 \mu_{\chi p}^2/ (\pi m_{Z'}^4)$) for various values of the mass splitting $\delta$, using electron recoils at Super-K, from TXS 0506+056 (left) and NGC 1068 (right). We include DIS effects at the AGN source. All plots assume that the gauge coupling of the new mediator to electrons and protons is equal. For comparison, we show in shaded red existing bounds from direct detection and cosmological probes in the limit $\delta=0$ (see main text for details). Importantly, these bounds do not apply for the mass splittings considered in the figure ($\delta \gtrsim 0.01$ GeV). We also show in shaded gray color the excluded region for a combination of beam-dump and collider searches, with $\delta=0.1m_{\chi_1}$ and $m_{Z^{\prime}}=3m_{\chi}$ \cite{Izaguirre_2016, Mongillo:2023hbs}. Finally, we show some representative values of the DM-proton scattering cross section that can yield the relic for inelastic DM with the parameters considered, see Eq. \ref{eq:relic}. The upper plots correspond to a mediator mass of $m_{Z^{    \prime}}=0.05$ GeV, the middle plots to a mass of $m_{Z^{\prime}}=5$ GeV, and the lower plots to $m_{Z^{    \prime}}=500$ GeV. \label{fig:upper_limits_cross_section}}
\end{figure}

We can relax the previous assumptions on the mass splitting and relation between mediator mass and DM mass, in order to explore the strength of our constraints in other regions of the parameter space.
We show our constraints from Super-K on the non-relativistic DM-proton scattering cross section $\sigma_{\chi-p}$ (see Eq. \ref{eq:sigma_NR}) versus DM mass $m_{\chi}$, for various values of the mass splitting $\delta$ ($= 0.01,0.1,1,10\ \rm GeV$). We show results from $\chi_1$-electron scatterings at Super-K in Fig. \ref{fig:upper_limits_cross_section}. Complementary (weaker) results from proton recoils at Super-K are shown in the Appendix \ref{sec:proton_recoils_SK}. The left plots in Fig. \ref{fig:upper_limits_cross_section} correspond to limits from TXS 0506+056, and the right plots correspond to NGC 1068. Moreover, we also show limits for various values of the mediator mass $m_{Z^{\prime}}=0.05,5,500$ GeV.

We note from the plots that the mass splitting significantly impacts the strength of the constraints at high-DM masses, where larger mass splittings generically lead to less stringent constraints, since the scattering process is kinematically suppressed with respect to smaller values of the mass splitting.  We further notice that the constraints are more stringent for heavy mediators than for light mediators. This can be understood since the boosted DM fluxes are suppressed by the propagator in the DM-proton scattering cross section as  $\sim 1/(m_{Z^{\prime}}^2+2m_{\chi}T_{\chi_2}-\delta^2)^2$. The inverse suppression with the kinetic energy of the boosted DM $T_{\chi_2}$ manifests for lighter mediator masses over a wider range of kinetic energies than for heavy mediators, which leads to somewhat smaller boosted DM fluxes for heavy mediators for the energies relevant of Super-K, see Fig \ref{fig:boosted_DM_fluxes}.

In the limit $\delta \rightarrow 0$, our constraints can be compared to complementary bounds from direct detection, CR cooling in AGN, and cosmological probes. These bounds are shown as a shaded red region in the figure. In particular, for $m_{\chi} \lesssim 7 \times 10^{-4}$ GeV the leading bound arises from measurements of the number of relativistic species ($N_{\rm eff}$) at the time of Big Bang Nucleosynthesis (BBN) \cite{Depta:2019lbe, Giovanetti:2021izc, Krnjaic:2019dzc}. For $ 7 \times 10^{-4} \, \mathrm{GeV} \lesssim m_{\chi} \lesssim 3 \times 10^{-3} \, \mathrm{GeV}$, the leading bounds arise from CR cooling in NGC 1068 \cite{Herrera:2023nww, Gustafson:2024aom, Mishra:2025juk}. In the range $ 3 \times 10^{-3} \, \mathrm{GeV} \lesssim m_{\chi} \lesssim 0.2 \, \mathrm{GeV}$, the leading bounds arise from a search on direct detection of galactic CR boosted DM at Super-K \cite{Super-Kamiokande:2022ncz}. Finally, for $m_{\chi} \gtrsim 0.2$ GeV, traditional direct detection searches of galactic halo DM sets the most stringent bounds \cite{CRESST:2019jnq, LZ:2022lsv, Billard:2021uyg}.

The constraints from CR boosted DM from AGN can overcome previous bounds in the range $ 10^{-3}$ GeV $\lesssim m_{\chi} \lesssim 1$ GeV in the limit $\delta \rightarrow 0$ and $m_{Z^{\prime}} \gtrsim 0.05$ GeV, further testing large mass splittings. In the presence of sufficiently large mass splittings ($ \delta \gtrsim$ 400 keV), direct detection constraints do not apply. In this regime, inelastic DM is better probed in beam dumps and colliders \cite{Izaguirre_2016, Mongillo:2023hbs, Foguel:2024lca}. In Fig.~\ref{fig:upper_limits_cross_section}, the shaded gray region represents the combined exclusion limits from several experiments. 
For $10^{-3}~\mathrm{GeV} \lesssim m_{\chi} \lesssim 10^{-2}~\mathrm{GeV}$, the strongest bound is set by NA64~\cite{Mongillo:2023hbs}, where the $Z'$ is produced via dark bremsstrahlung and undergoes semi-visible decays, often with displaced visible products. 
In the range $10^{-2}~\mathrm{GeV} \lesssim m_{\chi} \lesssim 0.2~\mathrm{GeV}$, comparable limits arise from LSND~\cite{Izaguirre:2017bqb}, CHARM~\cite{Gninenko:2012eq}, and NuCal~\cite{Tsai:2019buq}, where the $Z'$ can be produced not only through dark bremsstrahlung but also via meson decays such as $\pi^0 \to \gamma Z'$, and the visible decay products of the heavier dark-sector state are typically displaced. 
For $m_{\chi} \gtrsim 0.2~\mathrm{GeV}$, the most stringent constraints come from collider searches, where the $Z'$ is produced in $e^+e^-$ or $pp$ collisions and decays semi-visibly. 
These include BaBar, LEP, and LHC results~\cite{Izaguirre_2016}: BaBar and LEP bounds stem from mono-photon plus missing-energy searches and invisible $Z$-width measurements, while LHC limits are derived from mono-jet/photon signatures and Drell--Yan--like processes, with sensitivity to both prompt and displaced visible decays.

We can also compute the relic abundance for non-zero mass splittings in our set-up. In the common limits $m_{\chi_1} \gg 2m_e$, $m_{Z^{\prime}} > 2m_{\chi_1}$ and $\delta < m_{\chi_1}$, the relic abundance is predominantly set by annihilations into electron-positron pairs, and the thermally-averaged cross section reads \cite{gonzález2021cosmology}
\begin{equation}\label{eq:relic}
\langle\sigma v\rangle \simeq\frac{\alpha g_{\chi}^2 g_e^2 ((2m_{\chi_1}+\delta)/2)^2}{(16\pi((2m_{\chi_1}+\delta)/2)^2-m_Z'^2)^2}.
\end{equation}
We confront our limits with the region of the parameter space that yields the DM relic abundance in the Universe via freeze-out ($\langle\sigma v\rangle \simeq 1.5 \times 10^{-36} \mathrm{~cm}^2$), for two different choices of the mass splitting ($\delta=0.01$ GeV in dashed purple, $\delta=10$ GeV in solid purple). We note however that the displayed thermal benchmarks are approximate and their validity is restricted to the regime $\delta \lesssim m_{\chi_1}$.

It can be noticed in Fig. \ref{fig:upper_limits_cross_section} that such freeze-out motivated region is fully tested for some mediator masses, while for others (\textit{e.g} $m_{Z^{\prime}}=5$ GeV) only the region around the resonance remains viable. Additionally to showing the benchmark values of the DM relic abundance, we also shade in black color the region of parameter space that is excluded by perturbativity, $g_{p}g_{\chi}>16\pi^{2}$. We note that this consideration is not relevant for light mediators, but becomes important for heavy mediators. For instance, for $m_{Z^{\prime}}=500$ GeV, we find that our limits from CR boosted DM probe regions of parameter space which are already excluded by perturbativity.

The constraints from CR boosted DM in AGN will be further strengthened with the future Hyper-Kamiokande experiment \cite{Hyper-Kamiokande:2018ofw}. We estimate the sensitivity enhancement to be mild though, within one order of magnitude. A brief discussion can be found in the Appendix \ref{sec:Scaling}.
%
%
\section{\label{sec:conclusions}
Conclusions}
We have derived novel constraints on the DM-proton and DM-electron scattering cross sections from the consideration that CRs in NGC 1068 and TXS 0506+056 can scatter off the surrounding DM in the vicinity of their supermassive black holes (SMBHs), yielding a flux of boosted DM particles on Earth. This scenario remains viable even when there is a mass-splitting in the dark sector. High-energy neutrino and gamma-ray observations indicate that CRs are accelerated in the vicinity of the central black hole in these sources, with large luminosities. This, together with the expectation that the DM density is expected to be very large in these environments, induces a potentially measurable flux on Earth, compensating the dilution with extragalactic distances.

In particular, we have derived constraints from Super-K (and XENONnT in Appendix \ref{sec:nuclear_recoils_Xenon}). We considered a realistic choice for the CR acceleration radius in these sources and computed the CR flux using leptohadronic models informed by observations. Importantly, we considered a concrete model of DM-proton and DM-electron interactions consisting of a fermionic DM candidate and a vector mediator. In addition, we also allow for a mass splitting in the dark sector. The latter possibility introduces a novel complexity in the analysis and rich new phenomenology, further allowing to evade the stringent bounds from direct detection experiments on galactic gravitationally-bound DM. We included decays from the up-scattered heavier DM state into the lightest and a photon or an electron-positron pair, which affects non-trivially the corresponding boosted DM fluxes. Finally, we include deep-inelastic scattering (DIS) at the source and at the detector, which has significant implications on the ensuing limits, especially at Super-K.

We report progress on several fronts. First, we find that the boosted DM flux from NGC 1068 generically overcomes that from TXS 0506+056 when DIS is included, due to the expected larger DM density at the source, as inferred from its measured black hole masses. In addition, the CR luminosities that have been considered in previous work at TXS 0506+056 may only apply during flaring periods, much shorter than the running exposures at XENONnT and Super-K. NGC 1068 is expected to accelerate CRs steadily, in contrast. Second, we find limits on inelastic light DM scenarios that overcome complementary bounds from collider and beam dump experiments, and allow us to test uncharted parameter space of thermal DM. Third, we find constraints in the non-relativistic DM-proton scattering cross section that can overcome previous bounds from direct detection, galactic CR boosted DM, CR cooling in AGN and BBN/CMB in the mass range from $m_{\rm DM} \sim 1$ MeV to $m_{\rm DM} \sim 1$ GeV, probing  thermal DM models. When considering a mass splitting among the two DM states, the constraints from direct detection shut off, and our phenomenological probes previously uncharted parameter space over a range of DM masses and mass splitting spanning several orders of magnitude.

A caveat of our work (and previous works on CR boosted DM) concerns the expected DM self-scatterings at the source, which we neglected, but that could affect non-trivially the energy distribution of the boosted DM fluxes on Earth. We will address the impact of self-scatterings in depth in upcoming work. Another important caveat is related to the DM density profile and corresponding DM column density around the SMBH in AGNs. Even if DM spikes are not formed and sustained in time, an NFW profile would still yield a large column density, although $\sim$ 3 to 4 orders of magnitude weaker than it is customarily assuming a DM spike \cite{Herrera:2023nww, Gustafson:2024aom}.

Future observations of high-energy neutrinos at IceCube from other AGN may allow us to set more stringent constraints by accumulating further statistics and reducing the uncertainties on the inferred CR spectra at these sources. Independent observational methods to infer the DM distribution in these environments are also crucial \cite{Sharma:2025ynw}, to reduce the degeneracy with the DM-CR scattering cross section. These points, together with the upcoming improvements in direct detection experiments (like XLZD or Darkside-20k \cite{XLZD:2024nsu, DarkSide-20k:2017zyg}) and Hyper-Kamiokande \cite{Hyper-Kamiokande:2018ofw}, will allow us to dive into thermal light/and or inelastic DM models further, hopefully leading us to even stronger constraints or better -- a detection.

\section*{Acknowledgments}
A.G., G.H., and I.M.S. are supported by the U.S. Department of Energy under the award number DE-SC0020262. M. M. and K. M. are supported by NSF Grant No. AST- 2108466. A.G. is partially supported by the World Premier International Research Center Initiative (WPI), MEXT, Japan. A.G. is grateful to QUP for hospitality during his visit. M.\,M. also acknowledges support from the Institute for Gravitation and the Cosmos (IGC) Postdoctoral Fellowship and the FermiForward Discovery Group, LLC under Contract No. 89243024CSC000002 with the U.S. Department of Energy, Office of Science, Office of High Energy Physics. The work of K.M. is also supported by the NSF Grants, No.~AST-2108467 and No.~AST-2308021, and KAKENHI No.~20H05852.

This material is based upon work supported by the U.S. Department of Energy, Office of Science, Office of Workforce Development for
Teachers and Scientists, Office of Science Graduate Student Research (SCGSR) program. The SCGSR program is administered by the
Oak Ridge Institute for Science and Education (ORISE) for the DOE. ORISE is managed by ORAU under contract number
DESC0014664. All opinions expressed in this paper are the author’s and do not necessarily reflect the policies and views of DOE,
ORAU, or ORISE. A.G. is a recipient of the SCGSR Award.

\bibliography{references}
\appendix
\section{Scaling \label{sec:Scaling}}
In this work, we have considered AGN with properties as specified in Table \ref{fig:table1}. Better future observations or models may alter our understanding of these values. Changes to $\Gamma_{B}$, $\theta_{\mathrm{LOS}}$, $\alpha_{p}$, and $\gamma'_{\max/\min,p}$ lead to non-trivial changes in the resulting event rate. Alterations to the other parameters, however, can be captured in a simple scaling relation. In this appendix, we will derive such a scaling.

We know that our rate will follow Eq. \ref{eq:diff_recoil_rate} (repeated below)

\begin{equation}
    \frac{dR}{dT_{i}} = N_{i}  \int dE_{1} \frac{d \Phi_1}{dE_1} \frac{d \sigma_{1}(E_{1})}{dT_{i}}.
\end{equation}

From Eq. \ref{eq-chi1-flux}, we know that the flux scales as

\begin{equation}
    \frac{d \Phi_{\chi_1}}{dE_{\chi_{1}}} \sim \frac{\Sigma_{\chi}}{d^2} \Gamma_{p} \sigma
\end{equation}
where $\Sigma_{\chi}$ itself depends upon the black hole mass, acceleration region, and extent of the DM spike. We know that $\sigma \sim g_{\chi}^2 g_{i}^2$. Thus, our number of events at a detector $N_{ev}$ will scale as

\begin{equation}
    N_{ev} \sim \frac{\Sigma_{\chi}}{d^2} L_{p} g_{\chi}^{4} g_{i}^{4} \mathcal{E},
\end{equation}
where $\mathcal{E}$ is the detector exposure. Therefore, if a number of events $N_{des}$ is required by an experiment to set an exclusion, limits on the coupling go as

\begin{equation}
    g_{\chi} g_{i} \sim \bigg( \frac{N_{des} d^2}{\Sigma_{\chi} L_{p} \mathcal{E}} \bigg)^{1/4}.
\end{equation}

Note, this assumes that the attenuation through the Earth is roughly the same. This is valid as long as the coupling does not exceed the ceiling value as calculated in Sec.~\ref{sec:Atten}.

As a quick exercise, let us estimate how our exclusions will change with data from Hyper-Kamiokande which is expected to have 10 times the exposure of Super-K. For energy/angular bins dominated by background, the desired number of events will scale as $\mathcal{E}^{1/2}$, while for background-free bins, the desired number of events is fixed. To that point we can estimate the future exclusions $(g_{\chi} g_{i})_{HK}$ to relate to our current exclusions $(g_{\chi}g_{i})_{SK}$ by

\begin{equation}
    (g_{\chi} g_{i})_{HK} = (g_{\chi} g_{i})_{SK} \times 
    \begin{cases}
        0.75 \mathrm{\, \, \, background \, dominated}
        \\
        0.56 \mathrm{\, \, \, background-free}
    \end{cases}
\end{equation}

\section{Different Cases of NGC 1068 \label{sec:NGC_Cases}}
As indicated in Fig.~\ref{fig:ngc1068_crlum} and Table~\ref{fig:table1}, there are multiple plausible cases for the proton spectrum at NGC 1068. Although our main text figures only show the results from Case 1 (C1), here we explore the impact from the other cases.

\begin{figure}[H]
    \includegraphics[width = 0.45 \textwidth]{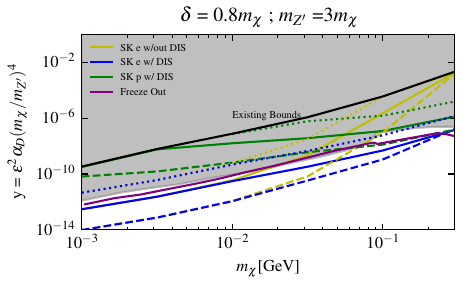}
    \includegraphics[width = 0.45 \textwidth]{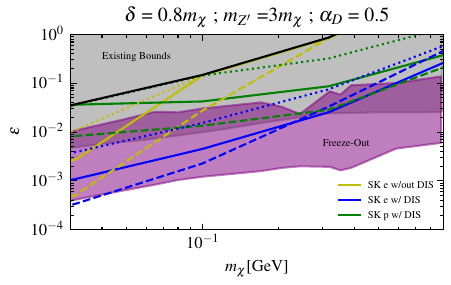}

    \caption{Exclusions similar to Fig. \ref{fig:inelastic_light_limits}, but for different cases of NGC 1068. Solid, dashed, and dotted lines indicate Cases 1, 2A, and 2B respectively.}
\end{figure}

We see that for the case of $\delta = 0.4 m_{DM}$ and $m_{Z'} = 10 m_{DM}$, Case 2A sets the strongest constraints, while Case 2B sets the weakest. Therefore, for these moderate mass splittings, we can treat Cases 2A and 2B as a range of possible constraints. For the larger mass splittings $\delta = 0.8 m_{DM}$ and electron scattering, Case 1 becomes the strongest for $m_{DM} \gtrsim 0.3$ GeV. This can be understood as large mass splittings requiring higher energies, and Case 1 has the highest luminosity as large energies. This effect is less extreme for proton scattering at Super-K, and Case 2A continues to set the strongest bounds.

\section{Upper limits from $\chi_2$ scatterings on Earth}

In the main text, we assumed that all upscattered $\chi_{2}$ decay long before reaching Earth. While this is a reasonable assumption considering the mass-splittings and distances in this work, it can still be elucidating to consider the impact of $\chi_{2}$ at Earth assuming no decays \footnote{In principle, one could consider a case where only a fraction of the DM decays to the ground state, but in this work we only consider the two extreme cases}.

First, we must find the flux of $\chi_{2}$ at Earth. This is relatively easy considering elastic scattering ($p + \chi_{1} \rightarrow p + \chi_{2}$). Since Earth is on-axis for the jet, we know that if the DM scatters at angle $\theta$, the original proton must have been oriented at angle $\theta$ to the jet (NGC 1068 does not have a jet, but its emission is isotropic, so the production angle is unimportant). Therefore, we can find our flux as
\begin{equation}
    \frac{d \Phi_{\chi_2}}{d T_{\chi_2}} = \frac{\Sigma_{\chi}}{m_{\chi} d^2} \int dT_{p} \frac{d\sigma}{dT_{\chi_2}} \frac{d \Gamma_{p}}{d T_{p} d \Omega} \bigg|_{\cos\theta(T_{p}, T_{\chi_2})},
    \label{eq:Chi2_sing_scat}
\end{equation}

where $\cos \theta(T_p, T_{\chi_{2}})$ is given by Eq. \ref{eq:Scattering_Angle}.

When DIS is included, the contribution to the $\chi_{2}$ flux is
\begin{equation}
    \frac{d \Phi_{\chi_2}}{d T_{\chi_2}} \bigg|_{\rm DIS} = \frac{\Sigma_{\chi}}{m_{\chi} d^2} \sum_{i} \int dT_{p} dx f_{i}(x,q^2) \frac{d\sigma_{q}(xs)}{dT_{\chi_2}} \frac{d \Gamma_{p}}{d T_{p} d \Omega} \bigg|_{\cos\theta(T_{q}, T_{\chi_2})}.
    \label{eq:Chi2_DIS}
\end{equation}

Then, the differential rate of elastic scattering can be determined as

\begin{equation}
    \int dE_{2}\frac{d \Phi_2}{dE_2} \frac{d \sigma_{2}(E_{2})}{dT_{i}} 
\end{equation}

with the cross section for $\chi_{2}$ scattering with point-particle $i$ is

\begin{equation}
    \frac{d \sigma_{2}(E_{2})}{d T_{i}} = \frac{g^2_{i} g^2_{\chi} m_{\chi}}{8 \pi m_{i} m_{Z}^4 ((\delta+m_{\chi})^2 - E_{2}^2)} \Big( -4 E^2_{2} m_{i} + 4 E_{2} m_{i} T_{i} - 2 \delta E_{2} (\delta + 2 m_{\chi})+2 m^2_{i} T_{i} + m_{i}(\delta^2 - 2 T^2_{i}) + T_{i}(\delta^2 + 2 m^2_{\chi} + 4 \delta m_{\chi}) \Big).
    \label{eq-dsigma2_dTSM}
\end{equation}
Form factors are included for protons and nuclei as needed.

\begin{figure}
    \centering
    \includegraphics[width = 0.49\textwidth]{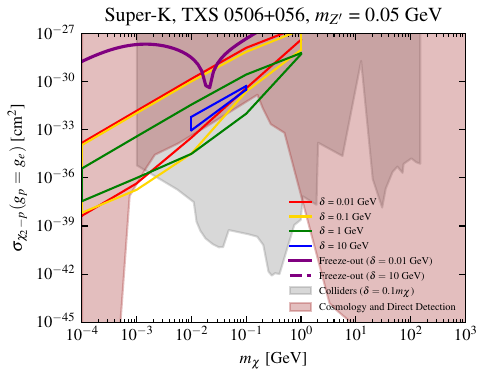}
    \includegraphics[width = 0.49\textwidth]{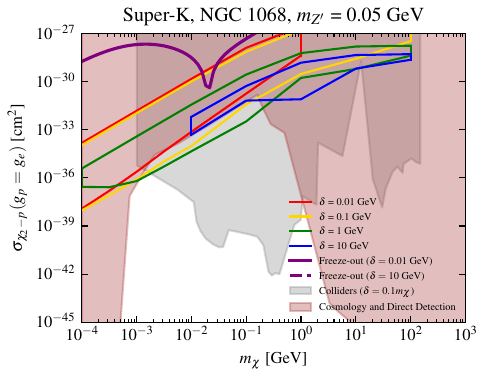}

    \includegraphics[width = 0.49\textwidth]{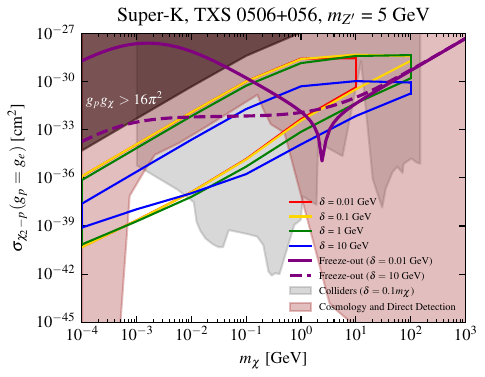}
    \includegraphics[width = 0.49\textwidth]{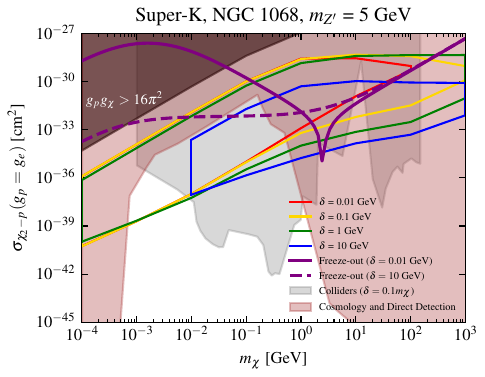}

    \includegraphics[width = 0.49\textwidth]{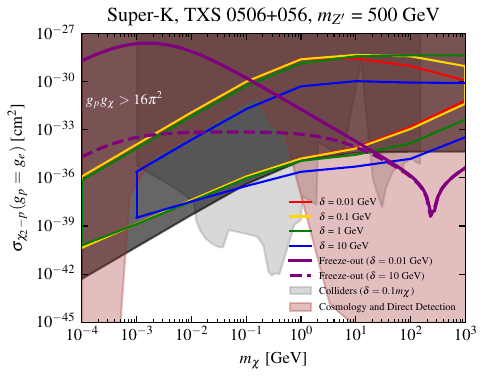}
    \includegraphics[width = 0.49\textwidth]{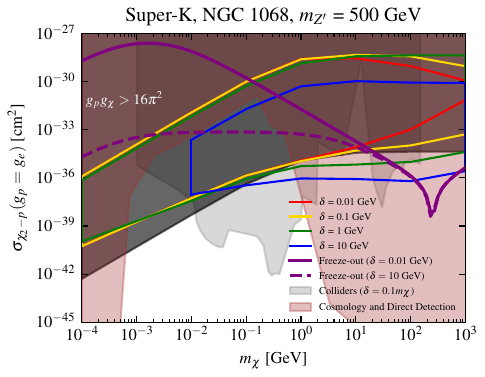}
    
    \caption{\justifying Same plots as \ref{fig:upper_limits_cross_section}, except neglecting decays so that only boosted $\chi_{2}$ arrives at Earth. In this case, there is no threshold energy for scattering at the Super-K detector. \label{fig:upper_limits_cross_section_chi2}}
\end{figure}

For DIS, we can compute

\begin{equation}
    \sigma_{2, \rm DIS}(s) = \sum_{i} \int dt dx f_{i}(x,-t) \frac{d \sigma_{2}(xs)}{dt}.
\end{equation}
where once again $\frac{d \sigma(s)}{dt} = \frac{1}{2m_{i}} \frac{d \sigma}{dT_{i}}$ and $T_{i} \rightarrow \frac{-t}{2m_{i}}$. We let $E_{2} \rightarrow \frac{1}{2m_{i}}\big( s - (m_{\chi} + \delta)^2 - m_{i}^2 \big)$. We show the resulting constraints in Fig. \ref{fig:upper_limits_cross_section_chi2}.

\section{Upper limits from proton recoils at Super-K}\label{sec:proton_recoils_SK}

As mentioned in the main text, we also derived upper limits on the DM-proton scattering cross section at Super-K from proton recoil data. This scenario, where the DM is boosted by CR protons at the source, and leaves signatures by proton recoils at the detector, allows us to connect our results with models where the DM is leptophobic, or couples predominantly to quarks. The upper limits are shown in Fig. \ref{fig:upper_limits_cross_section_protons_SK}, for TXS 0506+056 and NGC 1068 CR model C1 as an example. These limits are weaker than those obtained from electron recoils, but can still be leading in some regions of parameter space. Importantly, if the inelastic DM relic abundance is set by annihilations into protons, we find that thermal freeze-out scenarios are severely constrained.

\begin{figure}
    \includegraphics[width = 0.49\textwidth]{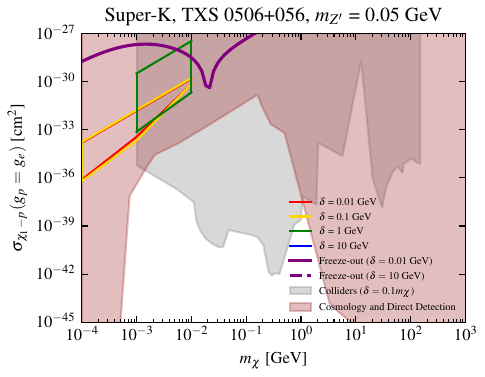}
    \includegraphics[width = 0.49\textwidth]{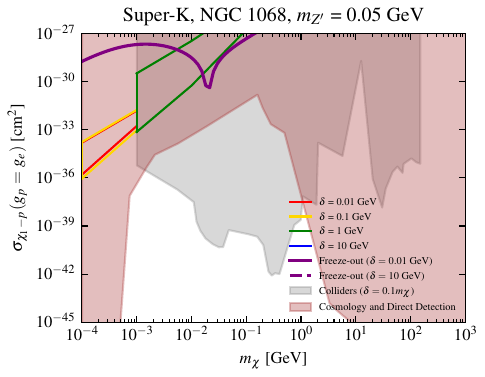}
    
    \includegraphics[width = 0.49\textwidth]{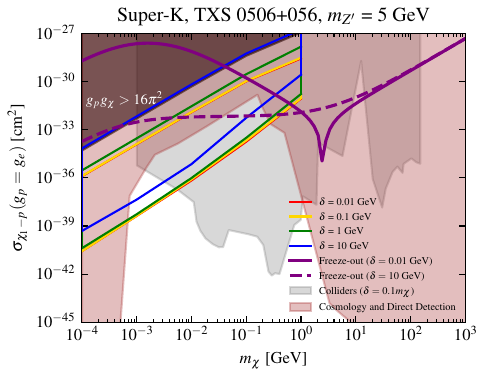}
    \includegraphics[width = 0.49\textwidth]{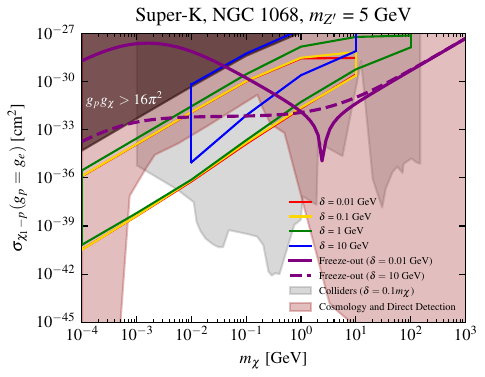}

    \includegraphics[width = 0.49\textwidth]{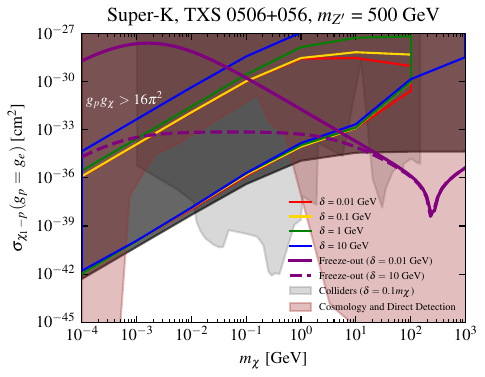}
    \includegraphics[width = 0.49\textwidth]{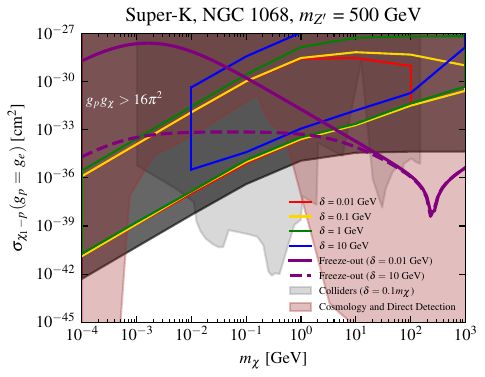}
    
    \caption{\justifying Same as Fig. \ref{fig:upper_limits_cross_section}, except considering DM-proton scatterings at Super-K. Here we show the constraints where DIS is considered at both the AGN and detector, as that leads to stronger constraints. \label{fig:upper_limits_cross_section_protons_SK}
    }
    \label{fig:upper_limits_cross_section_protons_SK}
\end{figure}

\section{Upper limits from nuclear recoils at XENONnT \label{sec:nuclear_recoils_Xenon}}

We also derive limits from nuclear recoils at the XENONnT experiment \cite{XENON:2023cxc}. Just as we did for Super-K, we can use Eq. \ref{eq:diff_recoil_rate} to calculate the differential recoil rate at XENONnT. Then, in order to find the total number of events in the energy window of the experiment, we calculate
\begin{equation}
N=\mathcal{E} \times \int_{T_A^{\min }}^{T_A^{\max }} d T_A \, \epsilon(T_A) \frac{d R}{d T_A} 
\label{eq:rate_xenon}
\end{equation}
where the minimum and maximum recoil energies of the xenon nuclei A are given by $T_A^{\rm min }=3.3$ keV and $T_A^{\rm max }=60.5$ keV. $\epsilon(T_A)$ denotes the efficiency of the experiment, which we take from \cite{XENON:2023cxc}. $\mathcal{E}$ refers to the exposure of the experiment, which is 1.09 tonne $\times$ yr. Finally, in order to derive an upper limit on the gauge couplings, mass splittings or non-relativistic cross section, we find the 90\% C.L upper limit on the number of signal events from a Poissonian likelihood as
\begin{equation}
\mathcal{L}\left(N_{\mathrm{obs}} \mid N_{\mathrm{sig}}+N_{\mathrm{bck}}\right)=\frac{\left(N_{\mathrm{sig}}+N_{\mathrm{bck}}\right)^{N_{\mathrm{bos}}}}{N_{\mathrm{obs}}!} e^{-\left(N_{\mathrm{sig}}+N_{\mathrm{bck}}\right)}
\end{equation}
where the number of observed events at XENONnT is $N_{\rm obs}=152$, and the expected number of background events is $N_{\rm bck}=149$. The 90$\%$ C.L upper limit on $N_{\rm sig}$ can then be obtained from the test statistic
\begin{equation}
2 \ln \left[\mathcal{L}_{\min }\left( N_{\mathrm{sig}}\right)\right]-2 \ln \left[\mathcal{L}\left( N_{\mathrm{sig}}\right)\right]=2.71
\end{equation}
getting $N_{\rm sig}^{90\% \rm C.L}<24.17$. Thus, we impose that the total rate induced by certain combinations of the parameters that we aim to constrain, confer Eq. \ref{eq:rate_xenon}, shall not exceed this upper limit. Under this prescription, we find the limits from XENONnT to be significantly weaker than those from Super-K.

\begin{figure}
    \includegraphics[width = 0.49\textwidth]{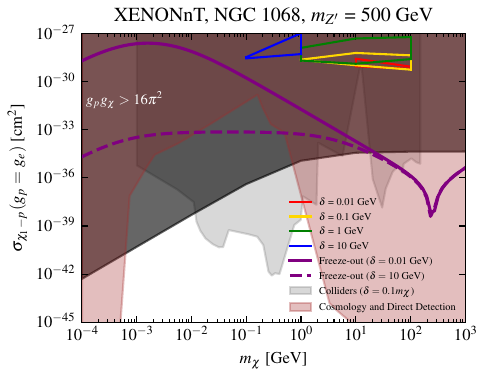}
    \caption{\justifying Upper limits on the non-relativistic scattering cross section of $\chi_1-p$, from nuclear recoils at XENONnT considering NGC 1068 as a source. We do not include other parameters that make panels in Figures \ref{fig:upper_limits_cross_section} and \ref{fig:upper_limits_cross_section_protons_SK}, as for those values, the flux experiences too significant an attenuation to set any constraints.
    }
    \label{fig:upper_limits_cross_section_protons_XENON}
\end{figure}

Our results are shown in Fig.~\ref{fig:upper_limits_cross_section_protons_XENON}. Depending on the values of the mass splitting, mediator mass, and $\chi_1$ mass, the flux may survive the attenuation in the atmosphere and reach the detector in some instances, but in most regions of parameter space the DM does not reach the detector (see Sec. \ref{sec:Atten} for more details). 

\end{document}